\documentclass[acmsmall]{acmart}

\setcopyright{cc}
\setcctype{by}
\acmDOI{10.1145/3832099}
\acmYear{2026}
\acmJournal{PACMSE}
\acmVolume{3}
\acmNumber{ISSTA}
\acmArticle{ISSTA008}
\acmMonth{10}
\acmSubmissionID{issta26main-p73-p}
\received{2026-01-30}
\received[accepted]{2026-06-25}

\usepackage{makecell}
\usepackage{xltabular}
\usepackage{balance}

\usepackage{tcolorbox}
\newtcolorbox[auto counter]{promptbox}[2][]{
    colback=white,
    colframe=black,
    colbacktitle=black,
    coltitle=white,
    fonttitle=\bfseries,
    sharp corners,
    boxrule=1pt,
    title={Prompt \thetcbcounter: #2}, 
    #1,                      
}
\definecolor{lightorange}{RGB}{255, 235, 205} 
\definecolor{lightpink}{RGB}{255, 230, 230}   
\definecolor{lightgreen}{RGB}{230, 255, 230}  
\definecolor{lightpurple}{RGB}{240, 230, 255} 

\usepackage{cleveref}

\newcommand{\tName}{\textsc{NL2Test}}

\begin{document}

\title{Industrial Practice of LLM-Based Test Case Carving and Assertion Generation (Experience Paper)}

\author{Haozhen You}
\orcid{0009-0000-1426-9041}
\affiliation{%
  \institution{Fudan University}
  \city{Shanghai}
  \country{China}
}
\email{hzyou23@m.fudan.edu.cn}

\author{Zhen Dong}
\authornote{Corresponding author.}
\orcid{0009-0009-1193-0696}
\affiliation{%
  \institution{Fudan University}
  \city{Shanghai}
  \country{China}
}
\email{zhendong@fudan.edu.cn}

\author{Jingjing Wang}
\orcid{0009-0009-0157-4050}
\affiliation{%
  \institution{ByteDance}
  \city{Shanghai}
  \country{China}
}
\email{wangjingjing.vector@bytedance.com}

\author{Qiang Li}
\orcid{0009-0003-4494-146X}
\affiliation{%
  \institution{ByteDance}
  \city{Shanghai}
  \country{China}
}
\email{liqiang.leo@bytedance.com}

\author{Xin Peng}
\orcid{0000-0003-3376-2581}
\affiliation{%
  \institution{Fudan University}
  \city{Shanghai}
  \country{China}
}
\email{pengxin@fudan.edu.cn}

\begin{abstract}
Enterprise regression testing for microservice systems is often constrained by incomplete or outdated documentation. In practice, QA engineers frequently rely on real execution traffic to reconstruct business scenarios; however, turning raw traffic into replayable regression tests with stable validation logic remains labor-intensive and error-prone.

This paper presents \tName, an end-to-end approach and tool that generates executable API regression tests from (i) a natural-language scenario description and (ii) a traffic capture recorded while executing the scenario. \tName\ addresses two coupled tasks: test case carving, which extracts a minimal replayable request sequence and reconstructs data dependencies so that dynamic values are bound from their responses rather than hard-coded; and assertion generation, which produces assertions aligned with business intent while avoiding non-deterministic fields and hallucinated paths. To improve reliability, \tName\ uses LLMs for semantic interpretation and constrained code synthesis, and uses deterministic algorithms for request filtering, dependency confirmation via value consistency, and assertion-path validation.

We evaluate \tName\ on 51 industrial regression scenarios extracted from a large consumer-facing Internet company. \tName\ achieves an exact-match rate of 82.4\% (42/51), and produces a functionally usable draft in 98.0\% (50/51) of scenarios when allowing minor post-edits. In a 9-month production deployment starting in March 2025, \tName\ generated 3,196 test cases with an overall code adoption rate of 85.4\%. These results indicate that traffic-grounded generation with deterministic guardrails can substantially reduce manual effort while improving regression automation in complex microservice environments.
\end{abstract}


\maketitle

\section{Introduction}

Enterprise software is increasingly organized as large-scale ecosystems of microservices, where requirements and APIs evolve rapidly and releases are frequent. In such settings, regression testing of cross-service business flows remains a substantial engineering burden. Although many testing activities have been automated, API regression tests for these flows are still largely constructed and maintained by QA engineers. Given a natural-language (NL) scenario description, engineers need to identify the relevant service endpoints, recover a replayable call sequence, bind dynamic values such as IDs and tokens across steps, and encode business expectations as executable checks~\cite{sathaye2023creating,andrzejewski2025automated}. This process is labor-intensive, error-prone, and difficult to scale: each API change or workflow update may require manual inspection of traces or logs, repair of data dependencies, and revision of assertions. As a result, generating and maintaining regression tests has become a practical bottleneck in industrial microservice systems.

Existing test-generation techniques provide only limited relief in this setting~\cite{sapozhnikov2024testspark,fakhoury2024llm,fraser2011evosuite}. Many techniques target a single module, class, or service, whereas real business scenarios often span multiple services. Others derive expected values from current executions, which can tie generated assertions to transient system behavior rather than intended business outcomes. Many approaches also assume isolated, stable, and resettable environments, an assumption that does not hold for many online services with shared state, dynamic content, and production-like dependencies~\cite{wang2024software,zhang2024trace,chen2023dynamic}. Consequently, generated tests may be difficult to replay reliably, brittle under data changes, or insufficient for checking the intended business behavior~\cite{guo2025comprehensive,jin2024llms}.

LLMs appear promising for reducing this manual burden because they can interpret natural-language scenarios and synthesize executable code. However, applying them directly to industrial microservice regression testing is difficult. The first challenge is selecting the business-relevant API sequence from noisy raw traffic. In practice, a captured trace may include background polling, authentication, logging, retries, monitoring requests, and calls from unrelated UI components. An LLM must distinguish the key APIs of the target business flow, preserve their order, and retain the dynamic data dependencies needed for replay. The second challenge is turning abstract business expectations into executable assertions. NL scenarios often describe outcomes such as successful creation, approval, or rule triggering, but do not specify which API response to inspect, which field or path to check, which operator to use, or which values are stable across executions. Consequently, an unconstrained single-pass NL-to-test approach may hallucinate endpoints, parameters, response fields, or assertions, leading to tests that are executable but irrelevant, flaky, or semantically incorrect~\cite{chen2024chatunitest,zhang2025exploring}.

This paper presents \tName, an end-to-end LLM-based tool deployed in an industrial microservice testing environment to assist QA engineers in generating API regression tests. Given a natural-language scenario description and raw traffic captured during scenario execution, \tName\ produces an executable test draft for engineer review and integration into existing regression suites. Rather than treating test generation as a single-step natural-language-to-code task, the paper reports the design choices required to make this agent-based workflow practical in this setting: carving the target business flow from noisy traffic, reconstructing dynamic data dependencies such as IDs and tokens for replay, and translating abstract natural-language expectations into concrete test assertions. A central focus is where LLMs should be used for semantic interpretation and where deterministic mechanisms are needed to validate generated artifacts and mitigate hallucinations, such as irrelevant API selection, invalid assertion paths, and incorrect variable references.

As an industrial experience report, this paper contributes lessons and evidence from building, deploying, and evaluating \tName. First, we characterize the practical obstacles that arise when applying LLM-assisted test generation to production microservice workflows, including underspecified NL descriptions, outdated documentation, noisy traffic captures, dynamic cross-request data dependencies, and unstable execution environments. Second, we present the traffic-grounded design of \tName, in which noisy traces are carved into replayable API sequences with runtime data bindings, and response-grounded assertion generation maps business intent to executable checks while filtering invalid paths and non-existent fields~\cite{kim2024leveraging}. Third, we report empirical results from both offline evaluation and deployment. On 51 industrial regression scenarios, \tName\ achieves an exact-match rate of 82.4\% and produces usable drafts for 50 scenarios. During the first nine months after its deployment in March 2025, \tName\ generated 3,196 regression tests, of which 2,730 were adopted by engineers, yielding an adoption rate of 85.4\%. These findings suggest that traffic-grounded LLM assistance can be integrated into production QA workflows, while also revealing limitations and lessons on where semantic LLM reasoning should be combined with deterministic validation.
\section{Traffic-Grounded Regression Test Generation in Industry}
\label{sec:problem}

\subsection{Scenarios, Traffic Captures, and Target Tests}
\label{subsec:setting}

In enterprise microservice systems, a user-visible business workflow is often implemented by a sequence of API calls across multiple services. In practice, however, engineers may not have access to complete source code, stable endpoint documentation, or up-to-date API specifications. When constructing regression tests for an existing business scenario, they often rely on two artifacts that are readily available in industrial testing workflows: a natural-language scenario description and a traffic capture recorded during one execution of that scenario.

These two artifacts provide complementary information. The natural-language scenario description specifies the business intent, user actions, and expected outcome. For example, it may describe that a user favorites a hashtag and then expects the hashtag to appear in the user's profile hashtag list. However, such a description does not specify the concrete backend APIs, request parameters, response fields, or data dependencies needed to implement the test. The traffic capture, in contrast, records the concrete HTTP/HTTPS requests and responses produced during the scenario execution. In Web systems, HAR is a common representation because it preserves request paths, methods, headers, cookies, request and response payloads, and timing information in a structured format~\cite{ran2024guardian,mascia2025microservices}. Therefore, traffic captures provide execution evidence that can ground API-level regression-test generation.

However, a traffic capture is not itself a regression test. It usually over-approximates the target scenario. A short user operation may trigger many requests, including static resource loading, background polling, preflight checks, retries, logging, recommendation refreshes, and unrelated UI-triggered calls. Replaying the entire capture would produce long and brittle tests that are sensitive to irrelevant implementation details and nondeterministic background behavior. Conversely, relying only on the natural-language scenario is insufficient because it lacks executable API-level details. The key problem is therefore to combine the semantic guidance from the scenario with the execution evidence from the traffic capture.

The target output in this paper is an executable API regression test for the given scenario. Such a test should satisfy three requirements. First, it should contain a compact sequence of business-relevant API calls carved from the noisy traffic. Second, it should be replayable across executions: dynamic values such as object identifiers, cursors, tokens, and session-dependent parameters should be extracted from earlier responses and reused in later requests, rather than being blindly hard-coded from the recorded trace. Third, it should contain business-level assertions over concrete response fields, so that the generated test checks whether the user-visible expectation in the scenario is satisfied instead of merely checking HTTP status codes.

In current industrial practice, engineers perform this process manually~\cite{alshahwan2024observation,yandrapally2023carving}. They inspect the captured traffic, identify requests that correspond to scenario steps, remove irrelevant calls, recover cross-request data dependencies, and translate natural-language expectations into executable assertions. This process is feasible for simple traces, but becomes labor-intensive and error-prone in large microservice systems where a single user action may trigger many auxiliary requests and where key runtime values are generated dynamically. This motivates traffic-grounded regression-test generation: given a natural-language scenario description and a recorded traffic capture, automatically produce a compact, replayable, and assertion-rich API regression test~\cite{elbaum2008carving,deljouyi2023generating,feng2025get}.

\subsection{Running Example: Favorite Hashtag}
\label{subsec:example}

\begin{figure}[H]
    \centering
    \includegraphics[width=\linewidth]{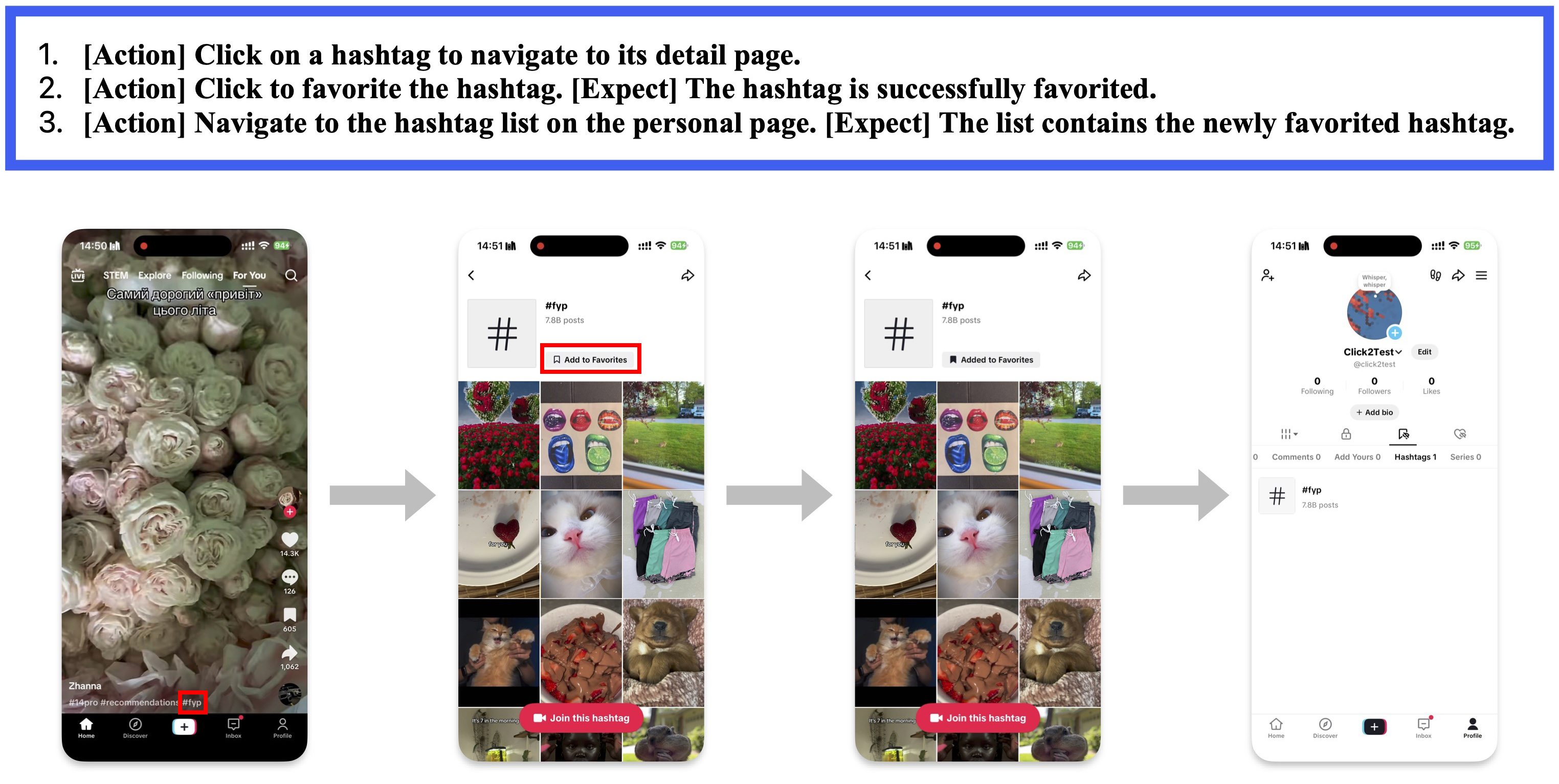}
    \caption{Example of traffic-grounded test carving and assertion generation.}
    \label{fig:example}
\end{figure}

Fig.~\ref{fig:example} shows a representative production scenario used as a running example in this paper. The scenario, titled \emph{Favorite Hashtag}, describes a common interaction in a content-consumption application: a user clicks a hashtag in the For You feed to navigate to its detail page, adds the hashtag to Favorites, and then checks whether the newly favorited hashtag appears in the user's Profile Hashtags list.

Although the scenario is concise from the user's perspective, the recorded traffic is much larger and noisier. Besides the requests that implement the hashtag-favoriting workflow, the capture may also include requests for resource loading, recommendation updates, logging, background refreshes, and other auxiliary interactions. A generated regression test should therefore not replay the whole capture. Instead, it should carve out the key API calls that realize the scenario, such as retrieving the selected hashtag from the feed or hashtag-detail response, sending the request that favorites the hashtag, and querying the Profile Hashtags list to validate the final state.

The carved sequence must also be made replayable. For example, the hashtag identifier observed in the recorded execution should not be hard-coded into the generated script. The test should extract the identifier, such as \texttt{hashtag\_id}, from an earlier response and reuse it in the later favorite request and in the final Profile Hashtags assertion. Without such dependency reconstruction, the generated script may pass only for the recorded execution and fail when runtime data changes in later executions.

Finally, the test needs business-level assertions. In this example, the generated assertions should check that the relevant feed or detail request succeeds, that the response contains a valid hashtag, that the favorite operation succeeds and marks the target hashtag as favorited, that the Profile Hashtags request succeeds, and that the returned Profile Hashtags list contains the same hashtag selected and favorited in the previous steps. These checks connect the natural-language expectation in the scenario to concrete fields in real API responses.

This running example illustrates the end-to-end task addressed in this paper. Given a natural-language scenario and a noisy traffic capture, the system must identify business-relevant requests, reconstruct data dependencies across requests, and generate executable assertions over response fields. \tName\ is designed for this setting: it performs traffic-grounded test carving and assertion generation to produce replayable API regression tests from industrial traffic captures.
\section{LLM-Based Regression Test Generation for Microservice Systems}
\label{sec:tname}

Our approach, \tName, generates regression tests for microservice systems from a natural-language business scenario and a traffic trace recorded during a reference execution. To handle noisy traces, runtime-dependent values, and meaningful test oracles, the agent-based workflow is organized into two stages: \emph{Test Script Carving} and \emph{Assertion Generation}. The first stage reconstructs a replayable business workflow by filtering irrelevant requests, matching scenario steps to request records, parameterizing runtime-dependent values such as tokens and object identifiers, and assembling the executable script. The second stage turns the replayable script into a regression test by deriving expected outcomes from the scenario, mapping them to concrete response fields, validating these fields against recorded responses, and inserting executable assertions into the script.

\subsection{Test Script Carving}
\label{subsec:test-script-extraction}
\begin{figure}[H]
    \centering
    \includegraphics[width=0.95\textwidth]{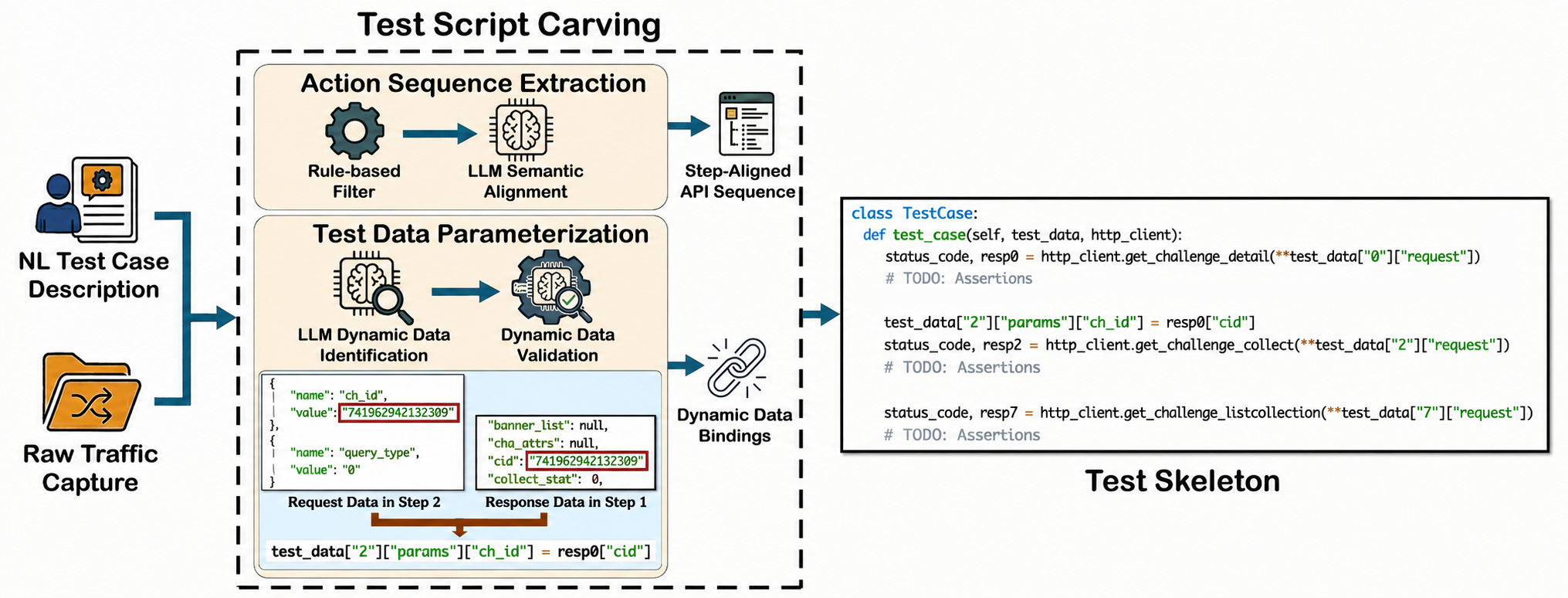}
    \caption{Overview of Test Script Carving.}
    \label{fig:carving_overview}
\end{figure}

Test script carving transforms a recorded traffic trace into a compact and executable API workflow. In our experience, traffic collected from microservice systems cannot be directly replayed as regression tests: it contains infrastructure requests, background service calls, and API calls unrelated to the target business scenario. Therefore, this stage extracts only the business-relevant trace and turns it into a runnable test skeleton. As shown in Fig.~\ref{fig:carving_overview}, \tName\ implements this stage as an agent-based workflow consisting of four coordinated steps: traffic filtering, LLM-based semantic alignment, dynamic data parameterization, and script assembly.

\subsubsection{Traffic Filtering.}

We first reduce the raw traffic through rule-based filtering before invoking the LLM. A practical lesson from our experience is that many irrelevant requests can be identified without semantic reasoning. We mainly filter the following categories of requests:

\begin{itemize}
    \item \textbf{Static resource requests.} These include JavaScript files, CSS files, images, fonts, icons, and other assets used for page rendering. They can usually be identified by path suffixes, content types, or cache-related headers.

    \item \textbf{Protocol-level requests.} These include browser- or runtime-generated requests such as CORS preflight \texttt{OPTIONS} calls. They are required for protocol negotiation, but do not represent user-level business actions.

    \item \textbf{Infrastructure and periodic requests.} These include heartbeat, keep-alive, logging, prefetching, background polling, and similar requests. They are often triggered by the application runtime rather than by the scenario step itself.
\end{itemize}

Sending all these requests to the LLM increases token cost and may distract the semantic-alignment agent from the API calls that actually implement the business workflow, making LLM-based semantic alignment less stable.

Our approach therefore applies conservative rule-based filtering based on request-level signals, including HTTP method, path suffix, content type, and repeated polling patterns. The filtering is intentionally conservative: if a request cannot be confidently classified as noise, it is kept for later semantic matching. This design choice reflects a trade-off we observed in practice. Removing a business-relevant request is difficult to recover from, whereas keeping a few extra candidates only increases the burden of the subsequent matching step.

\noindent\textbf{Rationale.}
Rule-based filtering is effective for removing structurally obvious noise. Within the agent-based workflow, this deterministic preprocessing step reduces the search space before semantic reasoning is invoked. LLMs are more useful after this reduction step, where the remaining task requires semantic understanding rather than simple pattern recognition.

\subsubsection{LLM-Based Semantic Alignment.}

After filtering, \tName\ uses a semantic-alignment agent powered by an LLM to perform semantic alignment between natural-language scenario steps and candidate endpoint records. Instead of asking the LLM to directly generate test code, we formulate this stage as a constrained prompt-based alignment task. Given the scenario description and the filtered traffic, the agent first infers the likely purpose of each request from its endpoint path, request parameters, and response fields, and then maps scenario steps to the corresponding request records.

This intermediate step-to-endpoint mapping is important in practice. In microservice traces, different requests may have similar paths or parameter names but correspond to different business actions. Directly generating a script from the scenario and traffic made such mistakes hard to inspect. By requiring the semantic-alignment agent to output an explicit step-to-endpoint mapping, engineers can check whether the extracted workflow is correct before the script is assembled.

The alignment prompt imposes several constraints. Each scenario step can be aligned with at most one endpoint record, and each endpoint record can be used for at most one step. The aligned endpoint records must preserve their chronological order in the original trace. The agent is also allowed to leave a step unmatched, since some scenario descriptions describe UI observations or navigation actions that do not trigger business API calls. Prompt~\ref{prompt:action-seq} shows the prompt used in this stage. Its output includes the request method, endpoint path, request identifier, and inferred request purpose, which together serve as the basis for script generation.

We further use few-shot examples to stabilize both the alignment pattern and the output format. The examples are selected from manually verified development cases and are excluded from the evaluation benchmark. Rather than selecting examples by business domain, we choose them by decision pattern: straightforward step-to-endpoint alignment, steps with no corresponding endpoint, and cases where temporal order is needed to distinguish similar endpoints. This strategy helps the agent follow the alignment constraints and produce structured output without overfitting to product-specific endpoint names or business objects.

\begin{promptbox}[label=prompt:action-seq]{Prompt for Step Alignment}
\vspace{0.2em}
\footnotesize
\colorbox{lightpink}{\textbf{Input Data}} \\
\textbf{Test Steps:} \colorbox{lightpurple}{\texttt{\{\{TEST\_DESC\}\}}} \\
\textbf{Filtered Traffic Data:} \colorbox{lightpurple}{\texttt{\{\{REQ\_LIST\}\}}} .

\vspace{0.2em}
\colorbox{lightorange}{\textbf{Task List}} \\
\textbf{Task 1:} Infer the likely purpose of each candidate endpoint from its path and request/response content.\\
\textbf{Task 2:} Match each step to at most one endpoint, and each endpoint to at most one step.

\vspace{0.2em}
\colorbox{lightorange}{\textbf{Constraint}} \\
1. Preserve the original step order; Do not match a later step to an earlier request.\\
2. Do not force a match when no business endpoint is triggered.

\vspace{0.2em}
\colorbox{lightgreen}{\textbf{Output:}} A JSON object containing step--endpoint mappings and endpoint purposes.

\vspace{0.2em}
\colorbox{lightorange}{\textbf{Few-shot Chain-of-Thought Example}} \\
(Same overall structure as above, not shown extensively here)

\end{promptbox}

\noindent\textbf{Rationale.}
Asking the agent to produce an inspectable intermediate representation is more reliable than asking it to generate test code directly. The step-to-endpoint mapping makes semantic errors visible early and allows deterministic procedures in subsequent workflow steps to assemble the executable script.

\subsubsection{Dynamic Data Parameterization.}

A correct endpoint sequence is still insufficient for regression testing. In practice, many generated scripts fail because they contain captured runtime values, such as user IDs, object IDs, content IDs, or session-dependent tokens. These values may be valid in the recorded execution but invalid when the test is replayed later. \tName\ addresses this problem through dynamic data parameterization, whose goal is to replace hard-coded request parameters with variables extracted from earlier responses.

\tName\ decomposes dynamic data parameterization into \emph{LLM Dynamic Data Identification} and \emph{Dynamic Data Validation}. In the agent-based workflow, the former acts as a semantic identification agent, while the latter combines deterministic validation with an LLM-based semantic check. This is motivated by the observation: directly asking the LLM to infer request dependencies is unreliable, because LLM may rely on field-name similarity and invent plausible but incorrect links. Instead, \tName\ first uses the LLM only to identify candidate dynamic fields, and then validates their dependencies using evidence from the recorded trace.

First, \tName\ uses the dynamic-data identification agent to identify request parameters that are likely to represent key business resources in the workflow. This step is semantic: the same business object may appear under different endpoint names or parameter conventions, and a single request may contain both control parameters and business identifiers. Thus, \tName\ uses this agent only to narrow the search space to resource-related request fields, rather than to directly infer dependencies.

Second, \tName\ validates whether the identified dynamic fields can be safely replaced by values extracted from previous responses. For a candidate parameter in request \(r_i\), the system retrieves its recorded value and searches previous responses \(r_j.response\), where \(j<i\), for the same value. If the value appears in a previous response, \tName\ records a candidate dependency from the response field to the later request parameter. This value-consistency check grounds the dependency in the observed trace and avoids relying only on field-name similarity.

Value consistency alone, however, can still be misleading, because the same value may appear in unrelated response fields. Therefore, \tName\ further applies a consistency check during Dynamic Data Validation. Given the source response path and the target request parameter path, the agent determines whether the two fields refer to the same business resource. A dependency is accepted only when both value consistency and semantic consistency hold. In cases with multiple valid candidates, \tName\ selects the earliest source that satisfies the semantic consistency check, following the chronological order of the recorded workflow.

The confirmed dependencies are used to rewrite the script. Hard-coded request values are replaced by variable references extracted from earlier responses. As a result, the generated test no longer depends on the concrete IDs observed in the original capture.

\noindent\textbf{Rationale.}
\tName\ uses LLM-powered agents only where semantic judgment is needed, while relying on value consistency to ground dependencies in the recorded trace.

\subsubsection{Script Assembly.}

Given the step-aligned endpoint sequence and the reconstructed dependency set, \tName\ completes the carving workflow by invoking a deterministic script-assembly step to assemble the executable test skeleton using templates. The templates encode stable script structure, including request construction, response parsing, variable extraction, and request sequencing. This avoids asking the LLM agents to generate the full script from scratch, reducing formatting errors and structural drift.

The resulting test skeleton contains the business-relevant API sequence and the necessary cross-step data flow. Assertion code is not generated at this stage; instead, \tName\ leaves explicit insertion points for the later assertion generation stage.

\subsection{Assertion Generation}
\label{subsec:assertion-generation}

After test script extraction, \tName\ obtains a replayable API workflow, but replayability alone does not make it a regression test. The workflow must also check whether the system behavior is correct. This is challenging because recorded responses contain many business-irrelevant fields, such as timestamps, trace identifiers, pagination cursors, and diagnostic messages, and treating them as expected outputs leads to brittle snapshot-style tests.

\begin{figure}[H]
    \centering
    \includegraphics[width=0.95\textwidth]{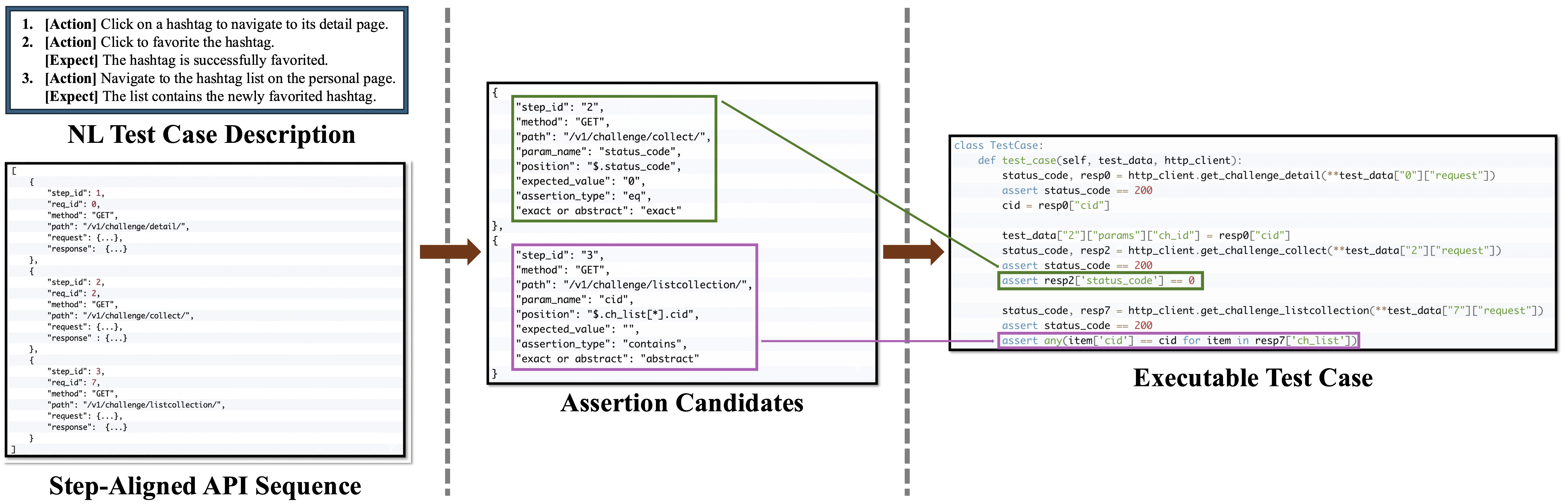}
    \caption{Overview of assertion generation and deterministic validation.}
    \label{fig:assertion_overview}
\end{figure}

Fig.~\ref{fig:assertion_overview} illustrates how \tName\ uses structured assertion candidates as an intermediate representation between the input scenario/traffic and the final executable test case. Each candidate records the target step, endpoint, response field, expected value, and assertion type, and is later translated into assertion code. Internally, \tName\ obtains these candidates through assertion intent extraction, deterministic assertion validation, and constrained assertion code generation. This design assigns bounded semantic decisions to LLM-powered agents, while leaving field grounding and script construction to deterministic procedures.

\subsubsection{Assertion Intent Extraction.}

\tName\ first identifies what should be verified after each business step. Instead of directly generating assertion code, an assertion-intent agent outputs structured \textbf{assertion candidates}. Each candidate records the step identifier, the matched endpoint, the target response field and its path, the assertion type, the expected value, and the expected-value form, i.e., \emph{exact} for fixed hard-coded values or \emph{abstract} for values resolved from dynamic variables during test execution.

Prompt~\ref{prompt:oracle-param} takes as input the natural-language test description, the step-aligned endpoint sequence, and compact response summaries. It asks the agent to infer business expectations from each scenario step and map them to observable fields in the corresponding endpoint response. This design avoids asking the agent to assert arbitrary fields from the whole trace.

\begin{promptbox}[label=prompt:oracle-param]{Assertion Intent Extraction}
\footnotesize
\vspace{0.2em}
\colorbox{lightpink}{\textbf{Input Data}} \\
\textbf{Test Step:} \colorbox{lightpurple}{\texttt{\{\{TEST\_DESC\}\}}} \\
\textbf{Request Sequence Set after Semantic Extraction:} \colorbox{lightpurple}{\texttt{\{\{REQ\_SEQ\}\}}} .

\vspace{0.2em}
\colorbox{lightorange}{\textbf{Task:}} Identify the business expectations implied by each step. Map each expectation to one or more response fields from the corresponding step.\\
\vspace{0.2em}
\colorbox{lightorange}{\textbf{Constraint}} \\
1. If the step contains \texttt{[expect]}, all mentioned fields must appear in the final assertion.\\
2. For state-change operations (e.g., like, follow, collect, upload, delete, update), verify that data was correctly added, removed, or updated. For list structures, use inclusion checks.\\
3. For every operation that succeeds, if a status code or equivalent field exists in the schema, a status code equality assertion must be generated.\\
4. Do not use runtime-only fields such as timestamps or log identifiers.

\vspace{0.2em}
\colorbox{lightgreen}{\textbf{Output:}} A JSON array of assertion descriptors.

\vspace{0.2em}
\colorbox{lightorange}{\textbf{Few-shot Chain-of-Thought Example}} \\
(Same overall structure as above, not shown extensively here)

\end{promptbox}

The prompt imposes several constraints. Explicit \texttt{[expect]} statements should be covered when they are observable from the response. For state-changing operations, such as like, follow, collect, upload, delete, or update, the candidate should check whether the relevant data has been added, removed, or updated. For list responses, containment-style assertions are preferred over full-list equality. Runtime-only or infrastructure fields, such as timestamps, trace identifiers, log identifiers, request durations, and diagnostic messages, are excluded. The agent may also output no Assertion Candidate for a step if no stable business-observable outcome is exposed.

We use few-shot examples to stabilize both the candidate format and the oracle-selection logic. The examples are selected from manually reviewed development cases and excluded from the evaluation benchmark. They are chosen by oracle pattern rather than by product domain, covering cases such as status-code equality, state-change confirmation, list containment, and abstract expected values.

The resulting assertion candidates separate semantic oracle inference from executable code generation. The assertion-intent agent proposes what should be checked, while later stages validate whether the fields exist, whether the expected values are stable, and how the assertions should be encoded.

\noindent\textbf{Rationale.}
Structured assertion candidates make the agent's oracle decisions inspectable and allow invalid or unstable checks to be filtered before code generation.

\subsubsection{Deterministic Assertion Validation.}

The assertion candidates produced by the assertion-intent agent may still contain invalid field paths, ambiguous field references, or unstable expected values. Therefore, \tName\ validates each Assertion Candidate against the recorded response before generating code. This step does not introduce new assertion intent; it only repairs, normalizes, or removes candidates proposed by the agent.

The two highlighted candidates in Fig.~\ref{fig:assertion_overview} show how this validation works in practice. In the first case, the agent proposes an \emph{exact} assertion over the response-body field \texttt{\$.status\_code} of the collect request. \tName\ checks that this path exists in the recorded response and retains the stable expected value \texttt{0}. The generated test therefore contains a business-level assertion over \texttt{resp2['status\_code']}, in addition to the generic HTTP-level assertion \texttt{status\_code == 200}. This distinction is important because the HTTP status code only indicates transport-level success, while the response-body status field captures the application-level outcome.

In the second case, the agent proposes an \emph{abstract} containment assertion over \texttt{\$.ch\_list[*].cid} for the list-collection request. \tName\ checks that the list field and its element field exist in the recorded response. Since the expected value is marked as abstract, the validator removes any hard-coded identifier produced by the agent and binds the assertion to the runtime variable \texttt{cid} obtained from an earlier step. Code generation then emits a containment check over the returned list, rather than a full-list equality assertion or a brittle hard-coded ID comparison.

For each Assertion Candidate, \tName\ validates the target field and expected-value form before code generation. It first checks whether the referenced response path exists. If the path is invalid, \tName\ repairs it by field-name search only when the response contains a unique matching field; otherwise, the candidate is discarded.

\tName\ then normalizes the expected value. Exact candidates are grounded to stable constants from the recorded response, whereas abstract candidates are treated as runtime-dependent values and are not hard-coded. Stable status fields, such as \texttt{status}, \texttt{status\_msg}, and \texttt{error\_msg}, are translated into exact equality assertions. Invalid or ambiguous candidates are removed to avoid brittle tests.

This step deliberately favors executable and stable assertions over aggressive assertion coverage. Removing an invalid Assertion Candidate may reduce the number of checks, but keeping a candidate that points to a non-existent field or hard-codes a dynamic value would generate a test that fails for the wrong reason.

\noindent\textbf{Rationale.}
Deterministic validation grounds the agent-generated candidates in recorded evidence, ensuring that every generated assertion references an actual response field and uses the correct expected-value form before code generation.

\subsubsection{Constrained Assertion Code Generation.}

After validation, \tName\ converts the refined assertion candidates into executable assertion statements and inserts them into the test skeleton. The code-generation agent's in this step is deliberately narrow. It is not allowed to invent new assertion targets, helper functions, response fields, or business logic. Instead, it receives the validated assertion descriptors, the relevant response structures, and the existing test skeleton, and generates code only for predefined insertion points.

This constrained generation is useful for assertions beyond simple templates, such as nested JSON checks, list-containment checks, and comparisons against variables extracted earlier in the workflow. Since assertion targets and expected-value forms have been validated, the agent only adapts the assertion code to the surrounding script context, including variable scope, indentation, response-access patterns, and the assertion library.

The final output is a complete executable regression test that combines the parameterized API workflow with the generated assertions. It can be replayed without invoking the agent-based workflow again. Regeneration is needed only when the business workflow, endpoint contract, response schema, or assertion semantics change materially.

\noindent\textbf{Rationale.}
The code-generation agent acts only as a bounded code translator: validated assertion candidates decide what to check, while the agent expresses those checks in the existing test skeleton.
\section{Evaluation}
\label{sec:experiments}

\subsection{Benchmark Construction}

To evaluate the generalization capability of \tName\ in a real-world enterprise environment\cite{sun2023presto}, we constructed a comprehensive benchmark dataset containing \textbf{51 distinct regression scenarios} collected from the production branches of five series at a large consumer-facing Internet company. Although all scenarios come from the same enterprise setting, these five series differ substantially in business style and technical complexity. This is because \tName\ requires natural-language descriptions of problematic test cases together with the corresponding server-side traffic, which are proprietary and generally inaccessible from other companies. As a result, we cannot conduct experiments across more enterprises or external environments, and instead evaluate \tName\ across diverse business lines within our own enterprise environment. The benchmark covers a range of application behaviors, from content consumption to content creation and transactional workflows. To maintain the conciseness of the main text, the detailed list of these scenarios and their specific attributes are provided in the Appendix~\ref{app:benchmark}. The benchmark covers the following domain-specific categories:

\begin{itemize}

    \item \textbf{C}-Series: Scenarios often require multiple assertions per step to check state changes and data integrity.

    \item \textbf{U}-Series: These scenarios tend to have higher noise in the recorded traffic due to repetitive user actions.

    \item \textbf{I}-Series: The interactions in these scenarios are more complex, often involving dynamic media and UI-related data manipulations .

    \item \textbf{M}-Series: These scenarios include workflows with dynamic parameter binding between steps.

    \item \textbf{S}-Series: Scenarios in this series focus on accuracy in handling dynamic queries, often dealing with complex check logic.

\end{itemize}

These scenarios are challenging for both carving and assertion generation. In production traffic, only a small fraction of captured requests corresponds to the actual business workflow. For example, in the representative \textit{U2} scenario, the system must identify only 2 valid business APIs from more than 300 raw requests. In addition, many scenarios require more than simple status checks and instead involve multiple business assertions over state changes or returned data.

\subsubsection{Ground Truth Formulation.}
For each scenario, we curated a triplet consisting of a natural-language test description, a raw traffic capture, and a ground-truth script. \textbf{The ground-truth scripts were manually written and reviewed by senior QA engineers} to ensure that they correctly reflect the intended business logic and handle cross-step dynamic dependencies. We use these scripts as the reference for evaluating the functional correctness of generated tests.

\subsubsection{Model Selection and Setup.}

For the experimental evaluation, we used a high-performance reasoning-oriented LLM, referred to as \textit{Enterprise-LLM} due to non-disclosure constraints on the underlying infrastructure. The model was deployed in a private cloud environment to ensure enterprise data isolation during generation. 

Importantly, \tName\ is an independent generation module rather than a system coupled to enterprise-specific infrastructure: except for benchmark construction and deployment on internal servers, its pipeline depends only on scenario descriptions, captured traffic, and LLM calls, making the reported effectiveness attributable to the generation method itself.

\subsection{RQ1: Effectiveness and Efficiency of Generation}

To answer whether \tName\ can generate functionally equivalent test scripts, we define a unified evaluation metric based on the generated code's execution and assertion status. An \textbf{Exact Match} case is one where the API sequence, dynamic parameter passing, and assertion logic are all strictly consistent with the ground truth, denoted by a checkmark ($\checkmark$). Cases that correctly identify the business workflow sequence but contain defects—such as missing or mismatched parameter dependencies or incomplete assertions—are classified as \textbf{Partial Match}, marked with an asterisk (*). Any case failing to identify the correct API sequence is deemed Incorrect.

Table~\ref{tab:metrics_details} presents the detailed results, with the column headers specifying the metrics reported for each scenario. \textit{Origin / Valid} contrasts the total number of raw captured requests against the number of valid business APIs, highlighting the noise level. \textit{Asserts} indicates the assertion complexity by counting the number of business rules in the ground truth. \textit{Defects (S/P/A)} reports the number of missing or incorrect workflow steps, parameter dependencies, and assertions, respectively. \textit{Tokens} and \textit{Time} quantify the computational cost in terms of LLM usage and total generation latency. \textit{Outcome} shows the final classification result.

\subsubsection{Effectiveness Analysis.}
The evaluation demonstrates that \tName\ is highly robust in producing production-ready scripts, achieving a \textbf{Total Correct Rate of 98.0\% (50/51)}, with \textbf{82.4\% (42/51) of these being Exact Match}. The high "Exact Match" rate is attributed to two key architectural choices tailored to different complexities. First, for scenarios with massive traffic volume (e.g., U1 with 157 requests), our Hybrid Denoising strategy—which rigorously filters structural noise via rules before applying the semantic-alignment agent—proved critical in isolating the valid API signal. Second, for scenarios requiring complex assertion (e.g., M12 with 7 assertions), the success is driven by our structured agent-based assertion generation workflow. By explicitly separating the identification of assertion entities from the analysis of their concrete values, the system can systematically construct multi-point assertions that match human-level rigor.

Regarding the "Partial Match" cases (15.6\%), the primary failure mode stems from the gap between text descriptions and human intuition. In Case S2 (Game Search), the text simply stated "Search input: Dota2" without defining specific checks. Consequently, the assertion-intent agent generated a generic assertion checking that the result list was non-empty, while the human ground truth additionally checked that the result title contained "Dota2". This discrepancy highlights that while the agent followed the provided instruction, it lacked the implicit domain habit of checking content relevance unless explicitly prompted. Similar assertion-side defects also appear in C9, C18, M2, M16, and M20. These cases represent valid but under-asserted scripts that are easy to patch. A secondary, less frequent cause is unmatched parameter names due to non-standard API definitions. In Case U3, the dynamic-data identification and validation steps failed to link a dependency because the API returned an ID with the obscure name `odidId' instead of a standard name like `userId'. Because the names looked completely unrelated, the agent cautiously avoided linking them.

The single Incorrect case (U2) reveals the tool's limitations when facing Repetitive Patterns combined with extreme noise. In this scenario, the user repeatedly invoked the same endpoint with varying parameters amidst 300+ background requests. The combination of an Extreme Signal-to-Noise Ratio and the similarity of repeated requests confused the semantic-alignment agent's step alignment, causing it to fail in identifying the correct sequence. The defect-count column further

{
\scriptsize
\begin{xltabular}{\textwidth}{c p{3.5cm} c c c c c c}
\caption{Detailed Evaluation Metrics including Accuracy, Token Usage, Execution Time, and Defect Counts.}
\label{tab:metrics_details}\\
\toprule
\textbf{ID} & \textbf{Description Title} & \textbf{\makecell{Origin /\\ Valid}} & \textbf{Asserts} & \textbf{\makecell{Tokens\\(Prompt / Compl.)}} & \textbf{Time(s)} & \textbf{\makecell{Defects\\(S/P/A)}} & \textbf{\makecell{Outcome}} \\
\midrule
\endfirsthead

\toprule
\textbf{ID} & \textbf{Description Title} & \textbf{\makecell{Origin /\\ Valid}} & \textbf{Asserts} & \textbf{\makecell{Tokens\\(Prompt / Compl.)}} & \textbf{Time(s)} & \textbf{\makecell{Defects\\(S/P/A)}} & \textbf{\makecell{Outcome}} \\
\midrule
\endhead

\midrule
\endfoot

C1 & Discover favorite hashtag & 8 / 3 & 4 & 25,457 / 12,095 & 442 & / & $\checkmark$ \\
C2 & View private profile videos & 3 / 3 & 4 & 49,094 / 5,491 & 204 & / & $\checkmark$ \\
C3 & Unfavorite video on profile & 3 / 2 & 3 & 16,654 / 10,169 & 349 & (0/1/0) & $\checkmark^*$ \\
C4 & Favorite video to new collection & 4 / 3 & 4 & 17,245 / 8,198 & 289 & / & $\checkmark$ \\
C5 & Create new video collection & 9 / 4 & 6 & 45,772 / 10,935 & 352 & / & $\checkmark$ \\
C6 & Delete video collection & 4 / 3 & 3 & 16,547 / 7,100 & 242 & / & $\checkmark$ \\
C7 & Cancel favorite hashtag & 8 / 3 & 5 & 23,646 / 7,485 & 230 & / & $\checkmark$ \\
C8 & View posts list video & 3 / 2 & 3 & 38,537 / 5,735 & 198 & / & $\checkmark$ \\
C9 & Set feed recommendation content & 4 / 3 & 4 & 18,179 / 6,888 & 228 & (0/0/1) & $\checkmark^*$ \\
C10 & Delete feed history instruction & 3 / 2 & 3 & 38,609 / 5,687 & 204 & / & $\checkmark$ \\
C11 & Delete feed instruction (Long press) & 4 / 2 & 3 & 16,649 / 4,139 & 143 & / & $\checkmark$ \\
C12 & Set new feed preference & 4 / 4 & 6 & 35,429 / 10,036 & 343 & / & $\checkmark$ \\
C13 & Unfavorite hashtag from list & 5 / 3 & 3 & 38,863 / 5,221 & 173 & / & $\checkmark$ \\
C14 & Update profile pronouns & 3 / 2 & 3 & 19,993 / 7,437 & 243 & / & $\checkmark$ \\
C15 & Switch to private account & 3 / 1 & 2 & 18,581 / 6,920 & 238 & / & $\checkmark$ \\
C16 & Switch to Business Account & 5 / 1 & 1 & 17,811 / 5,872 & 212 & / & $\checkmark$ \\
C17 & Reorder profile advanced features & 3 / 2 & 2 & 16,683 / 7,323 & 253 & / & $\checkmark$ \\
C18 & Link account & 4 / 1 & 2 & 16,829 / 4,222 & 144 & (0/0/1) & $\checkmark^*$ \\
\midrule
U1 & Like video (Logged-in) & 157 / 3 & 6 & 104,341 / 9,980 & 353 & / & $\checkmark$ \\
U2 & Login popup (Logged-out) & 307 / 2 & 2 & 97,643 / 6,908 & 259 & (1/0/2) &  \\
U3 & Follow another user & 152 / 2 & 4 & 29,298 / 5,332 & 194 & (0/1/0) & $\checkmark^*$ \\
U4 & Favorite video & 62 / 2 & 4 & 71,002 / 7,208 & 273 & / & $\checkmark$ \\
\midrule
I1 & Select video template & 2 / 2 & 4 & 16,358 / 3,700 & 132 & / & $\checkmark$ \\
I2 & View video effect details & 2 / 1 & 2 & 15,728 / 4,271 & 154 & / & $\checkmark$ \\
\midrule
M1 & Artist: Upload \& delete song & 113 / 1 & 2 & 82,356 / 10,572 & 372 & / & $\checkmark$ \\
M2 & Artist: Upload \& view releases & 77 / 2 & 5 & 39,066 / 6,587 & 245 & (0/0/2) & $\checkmark^*$ \\
M3 & Edit and save release draft & 165 / 2 & 3 & 108,567 / 7,689 & 311 & / & $\checkmark$ \\
M4 & Delete draft from 'My Releases' & 58 / 3 & 5 & 40,115 / 8,049 & 314 & / & $\checkmark$ \\
M5 & Download \& view AOP contract & 39 / 1 & 2 & 19,638 / 5,450 & 206 & / & $\checkmark$ \\
M6 & View artist insights & 51 / 2 & 4 & 36,860 / 4,613 & 184 & / & $\checkmark$ \\
M7 & Modify 'Under Review' song title & 87 / 3 & 4 & 59,782 / 8,649 & 618 & / & $\checkmark$ \\
M8 & View artist 7-day trend data & 13 / 2 & 3 & 20,561 / 9,718 & 333 & / & $\checkmark$ \\
M9 & Sort artist songs by posts & 10 / 3 & 6 & 21,437 / 6,197 & 215 & / & $\checkmark$ \\
M10 & Search and star artist & 11 / 3 & 4 & 19,900 / 6,826 & 238 & / & $\checkmark$ \\
M11 & Artist: Login \& view songs & 144 / 3 & 5 & 40,624 / 10,877 & 367 & / & $\checkmark$ \\
M12 & Artist: Modify 'Under Review' song & 163 / 4 & 7 & 69,856 / 6,673 & 235 & / & $\checkmark$ \\
M13 & Artist: Delete rejected song & 23 / 2 & 2 & 23,537 / 6,087 & 220 & / & $\checkmark$ \\
M14 & Label: Save draft \& delete & 158 / 3 & 5 & 46,081 / 7,419 & 261 & / & $\checkmark$ \\
M15 & Label: Upload artist avatar & 58 / 3 & 5 & 32,800 / 8,104 & 603 & / & $\checkmark$ \\
M16 & Label: Add new artist & 60 / 4 & 7 & 36,356 / 8,131 & 269 & (0/0/1) & $\checkmark^*$ \\
M17 & Exit song edit without save & 95 / 2 & 4 & 59,497 / 7,008 & 229 & / & $\checkmark$ \\
M18 & Artist: Login \& change avatar & 162 / 1 & 1 & 48,791 / 5,780 & 208 & / & $\checkmark$ \\
M19 & Creator joins WWA & 4 / 1 & 2 & 14,219 / 4,158 & 147 & / & $\checkmark$ \\
M20 & Creator submits WWA video & 18 / 3 & 5 & 26,134 / 11,603 & 392 & (0/0/1) & $\checkmark^*$ \\
M21 & Creator submits self-serve video & 10 / 1 & 2 & 16,382 / 4,286 & 164 & / & $\checkmark$ \\
M22 & View joined WWA campaigns & 14 / 2 & 3 & 20,025 / 5,328 & 202 & / & $\checkmark$ \\
M23 & Favorite WWA campaign & 16 / 3 & 5 & 26,269 / 7,859 & 280 & / & $\checkmark$ \\
M24 & View WWA bills & 14 / 2 & 3 & 19,853 / 8,443 & 312 & / & $\checkmark$ \\
M25 & Subscribe to WWA campaign & 14 / 2 & 3 & 20,391 / 5,630 & 203 & / & $\checkmark$ \\
\midrule
S1 & Unit conversion search & 1 / 1 & 4 & 36,335 / 8,700 & 308 & / & $\checkmark$ \\
S2 & Game card search & 1 / 1 & 3 & 16,739 / 5,886 & 222 & (0/0/1) & $\checkmark^*$ \\

\end{xltabular}
}

\noindent confirms that this failure is not merely an assertion-level issue, but involves an incorrect workflow step together with assertion omissions.

\subsubsection{Efficiency Analysis.}
In addition to effectiveness, we also analyze the efficiency of \tName\ in terms of generation latency and LLM token usage. Across all 51 scenarios, \tName\ takes an average of \textbf{265 seconds} to generate a test script. The average token consumption is \textbf{35.4k input tokens} and \textbf{7.2k output tokens} per case, resulting in an average total of \textbf{42.6k tokens}. Reporting input and output tokens separately provides a clearer view of the computational cost: the input side mainly reflects the cost of processing captured traffic, intermediate analysis results, and task descriptions, while the output side reflects the generated scripts and reasoning artifacts.

The token consumption varies substantially with the size of the captured traffic. For example, U1 and M3 require over 100k input tokens because they contain 157 and 165 raw requests, respectively, and the system needs to process large capture files during agent-based semantic alignment and workflow reconstruction. Execution time is also affected by the complexity of iterative assertion refinement, as scenarios with more complicated business logic or assertion construction may require additional interactions among the assertion-intent and code-generation agents.

Although an average latency of around four minutes is longer than instantaneous code completion, it remains practical in the context of integration-test generation. Manually writing, debugging, and validating a complex integration test often requires substantial engineering effort. Therefore, trading a few minutes of machine generation time for an executable test script can still provide meaningful productivity gains, especially for workflows involving noisy traffic, multi-step API dependencies, and non-trivial assertions.

\subsection{RQ2: Maintainability and Practical Quality in Deployment}

To evaluate the long-term maintainability and utility of the generated scripts, we introduced \tName\ into the deployment environment at the same company starting in \textbf{March 2025}. By the data cutoff on December 2025, the tool had successfully generated a total of \textbf{3,196 regression test cases}.

\subsubsection{Industrial Adoption Rate.}
The most rigorous measure of generated code quality is whether engineers are willing to merge it into the master codebase\cite{li2022automatically, alshahwan2024automated}. \tName\ achieved an impressive \textbf{Overall Adoption Rate of 85.4\% (2730/3196)}, indicating that the vast majority of generated scripts required little to no modification to meet deployment standards. This high acceptance rate confirms that the generated code is not only functionally correct but also adheres to the maintainability standards required by enterprise CI/CD pipelines.

\begin{figure*}[h]
    \centering
    \includegraphics[width=0.95\linewidth]{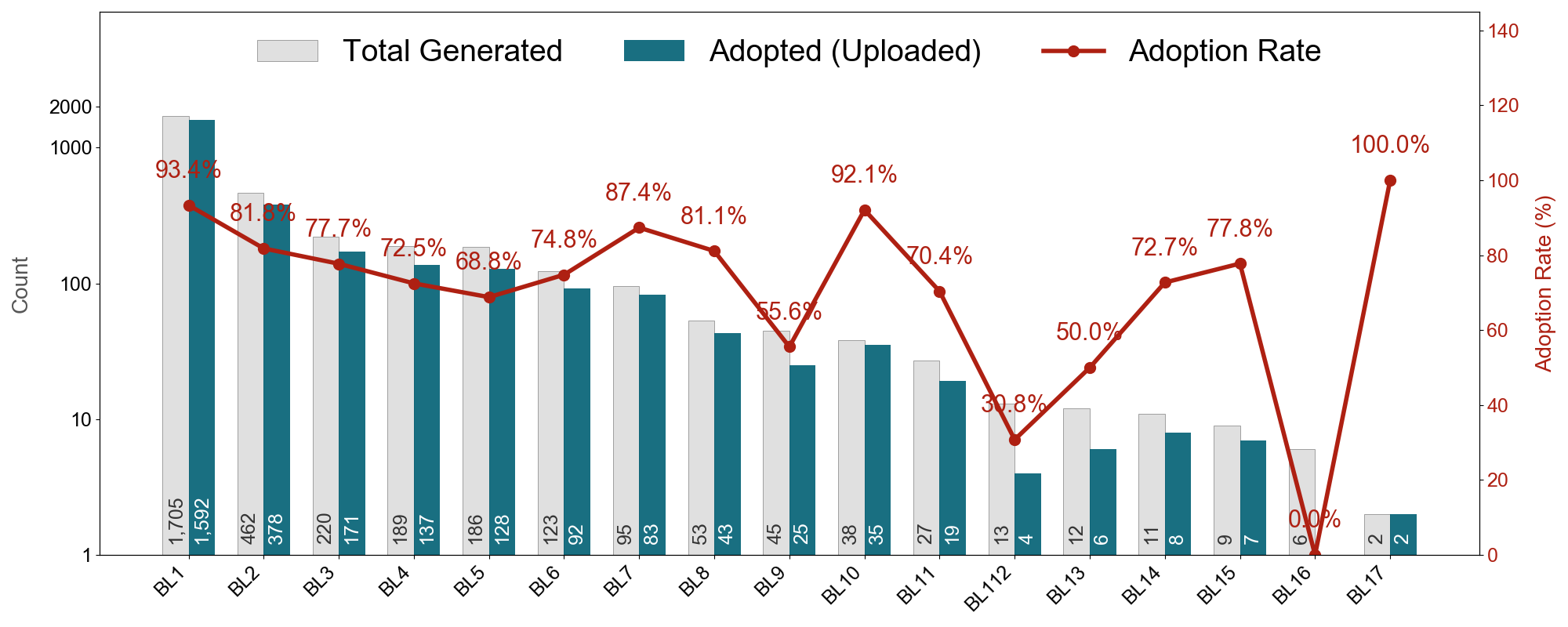}
    \caption{Adoption Rate of Generated Cases by Business Line.}
    \label{fig:adoption_rate}
\end{figure*}

Analysis of the business lines with lower adoption rates reveals specific environmental constraints that currently limit maintainability. The \textit{BL9} and \textit{BL13} business lines, for instance, recorded lower adoption rates of 55.6\% and 50.0\% respectively. This is primarily because these domains rely on highly dynamic, real-time content; while the tool correctly generates the request sequence, it struggles to generate precise, deterministic assertions for the rapidly changing data, forcing engineers to manually rewrite the assertion logic. Similarly, the \textit{BL12} and \textit{BL16} lines showed low adoption (30.8\% and 0.0\%  respectively). These domains rely heavily on non-standard binary protocols or asynchronous callbacks rather than standard HTTP interfaces. In such cases, the denoising module fails to isolate valid business steps from the protocol overhead, resulting in low-quality scripts that engineers opted to reject rather than repair.

\subsubsection{User and Expert Feedback.}
Complementing the quantitative adoption data, we conducted direct interviews with the leads of QA teams across 8 distinct business lines to gather their teams' overall feedback regarding the tool. Of these, \textbf{6 business lines provided explicitly positive evaluations}, emphasizing the tool's reliability and the high quality of the generated code.

\begin{table*}[htbp]
\centering
\caption{Summary of Expert Feedback and Key Metrics from Production Deployment.}
\label{tab:expert_feedback}
\scriptsize
\begin{tabular}{c p{10cm}}
\toprule
\textbf{Business Line} & \multicolumn{1}{c}{\textbf{Specific Feedback}} \\
\midrule
\textbf{BL1} & ``It used to take 3 hours to write automation for a single endpoint, but now, using the LLM to generate scenarios, doing multiple endpoints takes just 1 hour. Stability is hitting the mark too—between May 30th and June 5th, the success rate for scheduled tasks was 100\%.'' \\
\midrule
\textbf{BL2} & ``We saw some massive jumps in the data. P0-level API coverage is completely maxed out, going from 14.87\% straight to 100\%. Overall API coverage is now over halfway (at 50.68\%), and the server-side recall rate increased more than fourfold (from 6.36\% to 28.89\%). Basically, for all features released after April, we've achieved a 'shift-left' in API automation.'' \\
\midrule
\textbf{BL3} & ``Things are going well for us. After we added cases in the first phase, the coverage saw a huge boost. It jumped straight from the previous 29.01\% to 53.65\%—that's nearly double.'' \\
\midrule
\textbf{BL4} & ``Both the user experience and the generation results met our expectations. A real highlight is that it supports generating complex scenarios. Data-wise, we achieved a breakthrough in automation coverage, going from 0\% up to 9\% (specifically, bumping the test cases from 0 to 30).'' \\
\midrule
\textbf{BL5} & ``After using the tool, the full code coverage for P0 services shot up quickly, rising from 20\% to 30\% overall. At the microservice level, the biggest increase was around 40\%.'' \\
\midrule
\textbf{BL6} & ``Everyone feels it's really smooth to use, and the accuracy of the generated case statements is pretty high. No major issues there.'' \\
\bottomrule
\end{tabular}
\end{table*}

A summary of this feedback highlights that the tool significantly enhances workflow efficiency and test stability. According to the expert reports, the generated scripts exhibit high operational stability, with the \textit{BL1} team reporting a 100\% success rate for scheduled tasks running these cases. Furthermore, the usability is described as "smooth," with high accuracy in case statements, which directly translates to reduced maintenance overhead. The feedback also notes a tangible reduction in development effort, such as the time required per scenario dropping from 3 hours to 1 hour. The specific feedback from representative business lines is detailed in Table~\ref{tab:expert_feedback}. In addition, several teams reported that adopted \tName\ cases exposed real regressions after deployment, including functional logic errors, request failures or API connectivity issues, missing fields, and field-type mismatches.

\subsection{RQ3: Ease of Use and Organizational Adoption}

In this research question, we evaluate the tool's ease of use and its seamless integration into the daily workflow of the engineering team. We track the user journey and the volume of automation contribution over a 9-month deployment window to assess its vitality in a living environment.

\subsubsection{Continuous MAU Growth.} 
\begin{figure}[h]
    \centering
    \includegraphics[width=0.9\linewidth]{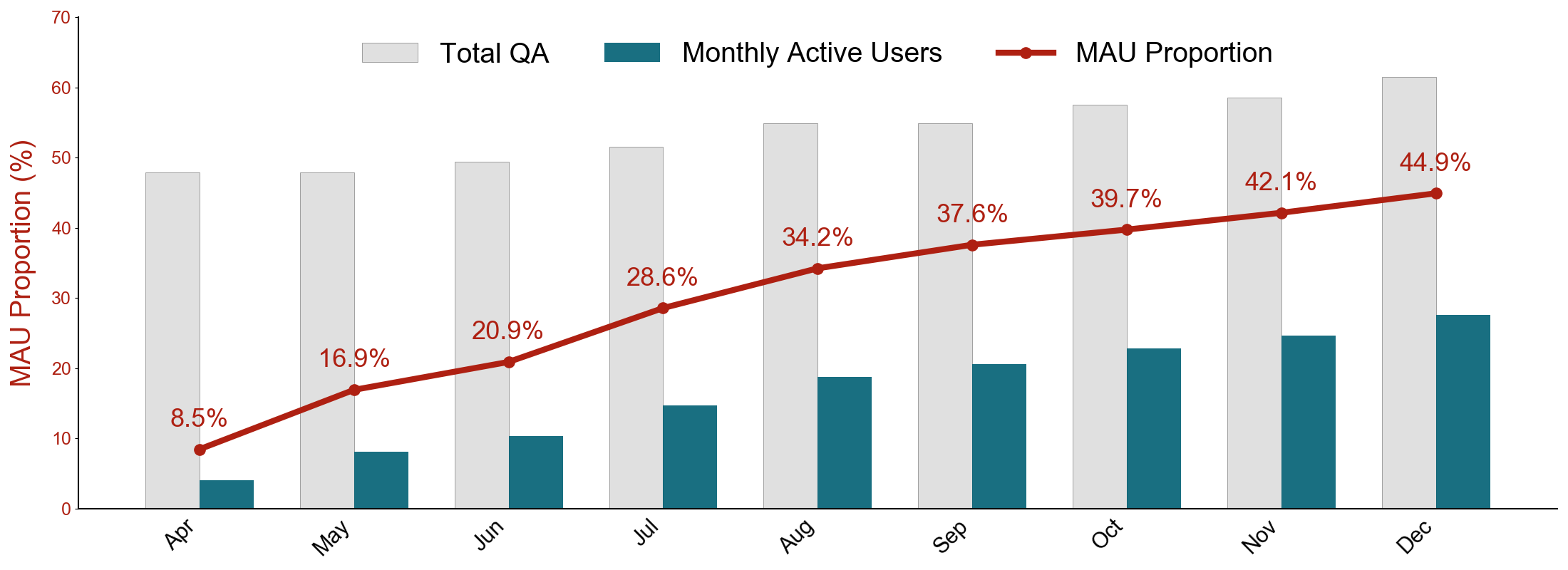}
    \caption{Monthly Active Users Proportion Trend.}
    \label{fig:mau_trend}
\end{figure}

Unlike many internal tools that hit mid-stage bottlenecks\cite{harman2025mutation, alshahwan2024enhancing}, \tName\ maintains an upward activity trend with strong penetration. Figure~\ref{fig:mau_trend} shows that active user growth significantly outpaced natural QA team expansion. Consequently, the monthly active user ratio rose steadily from \textbf{8.5} in April to \textbf{44.9} in December, indicating that nearly half of frontline engineers have integrated the tool into their daily workflows. This voluntary adoption, achieved without top-down mandates, suggests \tName\ drives organic growth by lowering technical barriers and addressing real engineering pain points.

\subsubsection{Proportion of LLM-Generated Test Cases.}
The tool successfully shifted from being an auxiliary utility to the dominant method for test creation, eventually accounting for \textbf{57\% of all new test cases} generated organization-wide during its peak usage period. The timeline of this adoption reveals a distinct "Trust Latency" phenomenon\cite{beller2023learning, saokar2023servicerouter}. In the initial six months, the proportion of generated cases remained modest as engineers tested the tool on non-critical features. However, following this validation period, the share of AI-generated cases surged, crossing the majority threshold in the fourth quarter. This shift suggests a fundamental change in the organizational workflow, where engineers have moved from manually authoring tests to reviewing and approving AI-generated scripts, validating the tool's ease of use and high trust level. Internal enterprise measurements show that, as of December 2025, \tName\ saves about \textbf{30 person-days every two weeks} on average compared with manual test authoring. This result demonstrates that the tool is not only technically effective, but also practically valuable at organizational scale.

\begin{figure}[h]
    \centering
    \includegraphics[width=0.95\linewidth]{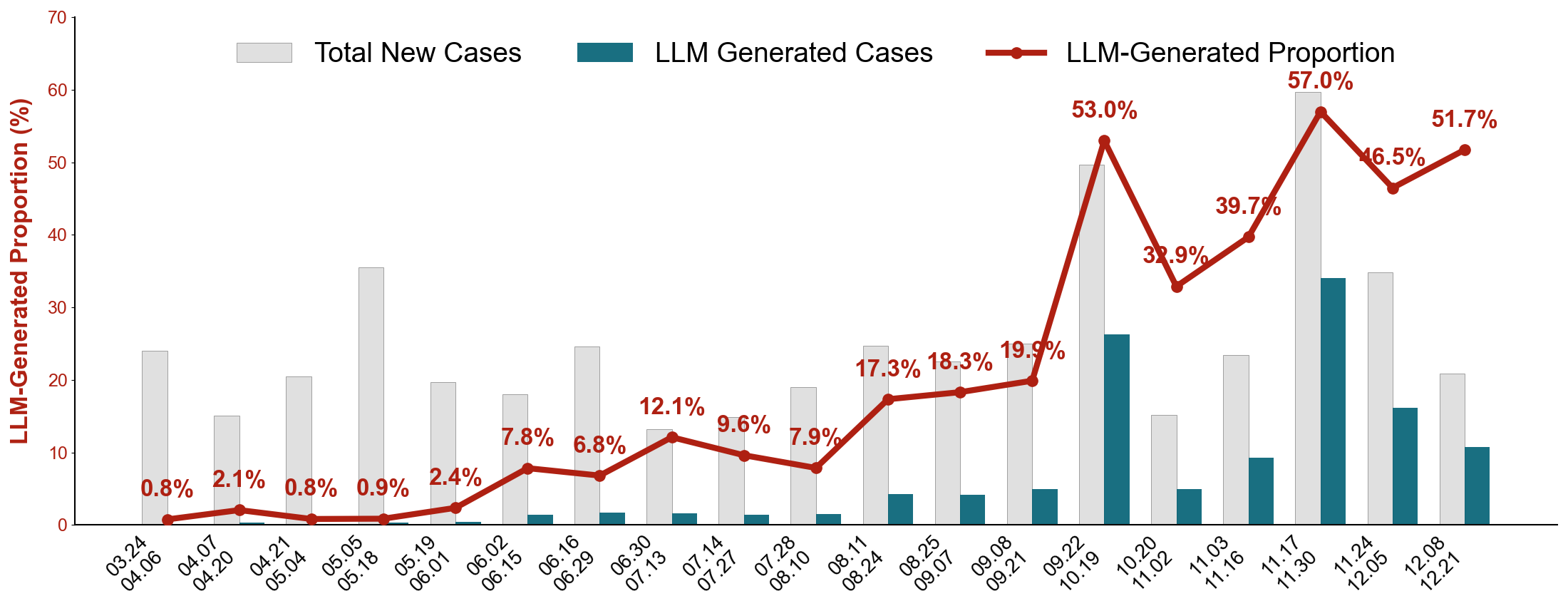}
    \caption{Case Penetration Trend (March-Dec).}
    \label{fig:case_trend}
\end{figure}

\section{Lessons Learned}
\label{sec:lessons-learned}

The development of \tName\ showed that reliable agent-based test generation depends less on larger prompts and more on limiting exact work delegated to LLM-powered agents. Many failures came from invalid intermediate artifacts, including non-existent fields, unstable assertions, incorrect bindings, and unexecutable code. We therefore adopted a constrained agent-based workflow: agents make bounded semantic decisions, while deterministic procedures validate, repair, or reject their outputs. These observations led to the following lessons.

\begin{enumerate}
    \item \textbf{Keep the Generation Pipeline Non-Blocking.} Early versions of \tName\ optimized each LLM stage independently, but we found that users preferred a lower-quality runnable draft to a higher-accuracy pipeline that stopped midway and required manual repair. Small structural errors could interrupt generation and make the system feel less like automation. We therefore made uninterrupted generation a design priority: uncertain LLM decisions are handled with schema checks, conservative repairs, or removal, so later stages can continue from partial but valid inputs.
    
    \item \textbf{Do not use the LLM for decisions that can be computed deterministically.} The LLM is useful for semantic interpretation, such as mapping business steps to requests or inferring expected outcomes, but unreliable for exact operations. In \tName, tasks such as field lookup, schema checking, value matching, request filtering, parameter substitution, and assertion validation are handled deterministically. LLM is used only when ambiguity or semantic reasoning is required.

    \item \textbf{Make each LLM task narrow enough to validate.} Large end-to-end generation tasks made errors difficult to locate and allowed early mistakes to propagate. We instead decomposed test generation into narrow tasks, such as request matching, dependency extraction, assertion intent extraction, and constrained code generation. Each model call produces a small structured artifact that can be validated before the next stage.

    \item \textbf{Give the model selected evidence, not the whole trace.} More context often introduced more noise. Raw traces contain static resources, polling requests, duplicated traffic, timestamps, trace identifiers, and diagnostic fields that can mislead the model. \tName\ therefore filters traces, summarizes response structures, highlights candidate fields, and excludes unstable runtime-only fields before invoking the LLM.

    \item \textbf{Constrain model outputs with allowed sets, not only natural-language instructions.} Natural-language instructions alone cannot prevent the model from inventing fields, variables, or helper functions. \tName\ provides explicit schemas and allowed sets, including valid request identifiers, candidate fields, assertion types, and existing variable names. Outputs outside these constraints are repaired only when a unique deterministic repair exists; otherwise, they are discarded.

    \item \textbf{Prefer stable partial assertions over rich but brittle assertions.} More assertions do not necessarily improve regression tests. Assertions over full responses, list order, timestamps, counters, or diagnostic messages often fail because of normal runtime variation. \tName\ therefore asserts only stable business-relevant fields and removes assertions whose targets are missing, ambiguous, or runtime-dependent.

\end{enumerate}

\section{Related Work}

Test carving has been studied as a way to derive focused tests from higher-level executions~\cite{elbaum2008carving}. Early work on carving and replaying differential unit tests shows that system executions can be transformed into more efficient unit-level regression tests. Recent industrial and mobile settings~\cite{dong2020time,guo2022detecting,sun2021understanding,cai2024reproducing,guo2025effectively} further validate this idea, e.g., ARTISAN carves Android GUI executions into JVM unit tests~\cite{gambi2023action}, and Meta’s TestGen generates regression tests from observed runtime values~\cite{alshahwan2024observation}. These works establish execution-grounded test generation, but they primarily operate at the program/object level rather than at the API-traffic level.

The closest work to ours is MicroTestCarver~\cite{deljouyi2023generating}, which carves E2E tests into understandable JUnit unit tests with meaningful scenarios and real test data. Our work shares its insight that E2E executions contain valuable business intent. However, MicroTestCarver targets method-level unit tests and relies on Java instrumentation, object reconstruction, and template-based generation. In contrast, our work targets API regression tests for microservice systems, using noisy HTTP traffic and a natural-language scenario, with key challenges in request selection, cross-request dependency recovery, and business assertion generation.

APICARV carves API tests from UI executions and infers OpenAPI specifications by recording and filtering browser-server traffic~\cite{yandrapally2023carving}. Unlike its specification- and coverage-oriented goal, our work focuses on scenario-based regression testing. Rather than infer general specifications, we reconstruct an executable business flow from noisy traffic, parameterize dynamic values, and generate assertions aligned with natural-language expectations.

LLM-based test generation and oracle generation are complementary to our work~\cite{fakhoury2024llm, wang2024software, molina2025test}. Systems such as ChatUniTest use LLMs to generate unit tests with validation and repair~\cite{chen2024chatunitest}, while TOGA studies neural test oracle generation under constrained oracle spaces~\cite{dinella2022toga}. Our work uses LLMs differently: rather than asking an LLM to directly generate a test or oracle, we ground generation in recorded traffic and constrain it with deterministic checks for request filtering, dependency validation, and assertion-path validation. Thus, our contribution is best positioned as traffic-grounded, LLM-assisted API test carving with business-oriented assertion synthesis, rather than general LLM-based unit test generation or pure API test carving.
\section{Threats to Validity}

The current results should be interpreted as evidence from one industrial deployment, and the external validity of \tName\ remains uncertain. Although the benchmark spans business lines and regression scenarios, all evidence still comes from a single large-scale enterprise environment.

\tName\ is also best suited to traffic-observable API regression workflows. It assumes that a scenario can be described in natural language and that a corresponding API traffic capture is available. This assumption is natural in our deployment setting, but its applicability to environments with incomplete captures or substantially different interaction patterns remains to be studied.
\section{Conclusion}

This paper presents \tName, an agent-based workflow for generating API regression tests from natural-language scenarios and recorded traffic. Combining LLM-powered reasoning with deterministic validation, \tName\ extracts replayable workflows, reconstructs dynamic dependencies, and generates stable business-oriented assertions. Industrial evaluation and deployment show that \tName\ reduces manual effort and produces maintainable regression tests. Future work will explore stronger oracles for dynamic states and deeper CI/CD integration.

\section{Data Availability}
The data is not publicly available due to the inclusion of proprietary service endpoints and potentially sensitive enterprise data.
\section*{Acknowledgements}
We thank the anonymous reviewers for insightful comments and suggestions. This work was supported by the National Key R\&D Program of China under Grant No. 2024YFB4505902.

\clearpage
\appendix
\section{Benchmark Scenarios}
\label{app:benchmark}
{
\scriptsize
\begin{xltabular}{\textwidth}{l X}
\caption{Benchmark Scenarios: Natural Language Test Descriptions} \label{tab:benchmark_desc}\\
\toprule
\textbf{ID} & \textbf{Natural Language (NL) Test Description} \\
\midrule
\endfirsthead

\multicolumn{2}{l}{Table ~\ref{tab:benchmark_desc} continued} \\
\toprule
\textbf{ID} & \textbf{Natural Language (NL) Test Description} \\
\midrule
\endhead

C1 & \textbf{[title] Discover favorite hashtag}
1. Enter Discover page; get hashtag list.
2. Select unfavorited hashtag (collect\_stat=0); click favorite [Expect] Success.
3. Check Profile hashtag list [Expect] New hashtag present. \\
\midrule
C2 & \textbf{[title] View private videos on profile page}
1. Enter Profile page.
2. Click Private tab [Expect] Private video list loads.
3. Click first video [Expect] Plays successfully. \\
\midrule
C3 & \textbf{[title] Unfavorite video}
1. Enter Profile page.
2. Click Favorites tab [Expect] Favorites list loads.
3. Select first video; click Favorite button [Expect] Unfavorited. \\
\midrule
C4 & \textbf{[title] Favorite video to a new collection}
1. Enter Feed Page.
2. Click Favorite button [Expect] Toast popup.
3. Click 'Manage' [Expect] Create Collection page.
4. Input valid name; click Save [Expect] Success. \\
\midrule
C5 & \textbf{[title] Create new collection}
1. Enter Collections page (list).
2. Click Create Collection [Expect] Creation page.
3. Input valid name; get video list.
4. Select first video; click Add Videos [Expect] Success. \\
\midrule
C6 & \textbf{[title] Delete collection}
1. Enter Collections list.
2. Select collection [Expect] Details page.
3. Click Delete Collection [Expect] Success. \\
\midrule
C7 & \textbf{[title] Cancel hashtags favorite}
1. Enter Favorited Hashtags list.
2. Select hashtag [Expect] Details page.
3. Click 'Added to Favorites' [Expect] Unfavorited.
4. Return to list [Expect] Hashtag removed. \\
\midrule
C8 & \textbf{[title] View posts list video}
1. Enter Posts list.
2. Click first video [Expect] Plays normally. \\
\midrule
C9 & \textbf{[title] Feed personalized recommendation content}
1. Open adjustment panel.
2. Select instruction; click Continue [Expect] Confirmation page.
3. Click 'Let's Go' [Expect] Success; feed refreshes. \\
\midrule
C10 & \textbf{[title] Feed delete historical personalized recommendation instruction}
1. Find video with 'Preference you set' tag.
2. Get instruction list [Expect] History page.
3. Delete first instruction [Expect] Success; feed refreshes. \\
\midrule
C11 & \textbf{[title] Feed delete historical personalized recommendation instruction}
1. Long press video; enter personalization page; get list.
2. Delete first instruction [Expect] History page updates. \\
\midrule
C12 & \textbf{[title] Feed personalized recommendation instruction}
1. Get feed video list.
2. Click 'Preference you set' tag on tagged video [Expect] History page.
3. Click '+Set New' [Expect] Adjustment panel.
4. Select instruction [Expect] Success; feed refreshes. \\
\midrule
C13 & \textbf{[title] Cancel hashtag favorite}
1. Enter Favorited Hashtag list.
2. Click hashtag [Expect] Details page.
3. Click Unfavorite [Expect] Success. \\
\midrule
C14 & \textbf{[title] Update pronoun}
1. Enter Profile.
2. Click Edit Profile.
3. Update Pronouns [Expect] status=200. \\
\midrule
C15 & \textbf{[title] Change account to private account}
1. Profile > Settings.
2. Enter Privacy.
3. Set to Private Account. \\
\midrule
C16 & \textbf{[title] Change account to BA account}
1. Profile > Settings.
2. Enter Account.
3. Switch to Business Account. \\
\midrule
C17 & \textbf{[title] Change display order of account advanced features}
1. Enter Profile.
2. Click Edit Profile.
3. Reorder features in Display Order. \\
\midrule
C18 & \textbf{[title] Account link}
1. Enter Profile.
2. Click Edit Profile.
3. Link other platform in 'Links' [Expect] status=200. \\
\midrule
U1 & \textbf{[title] Like video (logged-in user)}
1. Enter For You; get video list.
2. Like unliked video (digged=false) [Expect] Success.
3. Check Profile Likes list [Expect] Video present. \\
\midrule
U2 & \textbf{[title] Login pop-up (logged-out user)}
1. Enter Feed Page; get list.
2. Like unliked video [Expect] Login popup (3+ channels). \\
\midrule
U3 & \textbf{[title] Follow another user}
1. Enter Feed Page.
2. Follow unfollowed creator (follow\_status=0) [Expect] Success (status=1).
3. Enter Following page [Expect] Creator's videos visible. \\
\midrule
U4 & \textbf{[title] Favorite video}
1. Enter Feed Page; get list.
2. Favorite unfavorited video (collect\_stat=0) [Expect] Success.
3. Check Profile Favorites [Expect] List count >=1; video present. \\
\midrule
I1 & \textbf{[title] Select video template}
1. Click '+' (Shooting page).
2. Click Template (bottom right). \\
\midrule
I2 & \textbf{[title] View video effect}
1. Click video effect anchor [Expect] Effect details. \\
\midrule
M1 & \textbf{[title] Artist user selects single release, uploads \& deletes full song file}
1. Click Upload > Select 'Single Song'.
2. Enter track info page.
3. Add File (upload full song).
4. Delete song file [Expect] status=200. \\
\midrule
M2 & \textbf{[title] Single upload, click submit successfully then view all releases}
1. Fill track info; click Submit.
2. Click All Releases [Expect] Song list > 0. \\
\midrule
M3 & \textbf{[title] Secondary editing and saving of releases draft}
1. Releases > Drafts.
2. Edit one of drafts; Save. \\
\midrule
M4 & \textbf{[title] After saving the draft, delete the draft on the 'My Releases' page}
1. Click Upload > Single Song.
2. Fill partial info; click 'Save to My Drafts'.
3. Releases > Drafts.
4. Delete first draft. \\
\midrule
M5 & \textbf{[title] Contract management download and view AOP contract}
1. Click Avatar.
2. Select Contract Management.
3. Download first contract (status=200). \\
\midrule
M6 & \textbf{[title] View artist insights}
1. Releases > Artists.
2. View first artist details. \\
\midrule
M7 & \textbf{[title] Modify the title of an artist user's song 'Under Review'}
1. View All Releases.
2. View first song details.
3. Modify Title. \\
\midrule
M8 & \textbf{[title] View artist's data for the last 7 days}
1. Enter Artist Overview.
2. Select '7d' filter in Trends [Expect] Data shows 7d. \\
\midrule
M9 & \textbf{[title] View artist song list sorted by posts quantity}
1. Enter Song Performance.
2. Click 'Posts' header [Expect] Descending sort.
3. Click 'Posts' header [Expect] Ascending sort. \\
\midrule
M10 & \textbf{[title] On the artist list page, search by artist name and star an artist}
1. Enter Artist List.
2. Search 'Taylor'.
3. Click Subscribe on first result [Expect] Star=true. \\
\midrule
M11 & \textbf{[title] Artist user logs in and views songs}
1. Log in.
2. View uploaded songs [Expect] list > 0.
3. View album songs [Expect] rsp not none. \\
\midrule
M12 & \textbf{[title] Artist user modifies a song with 'Under Review' status}
1. View All Releases.
2. View unreviewed songs.
3. Edit unreviewed song; replace file.
4. Submit [Expect] Success (status=200). \\
\midrule
M13 & \textbf{[title] Artist user deletes a rejected song}
1. View All Releases.
2. Delete rejected song [Expect] Success (status=200). \\
\midrule
M14 & \textbf{[title] Label user saves a song to Drafts when uploading and then deletes it}
1. Click Upload > Enter title/lang.
2. Save to My Drafts [Expect] Success.
3. Delete draft. \\
\midrule
M15 & \textbf{[title] Label user uploads an avatar for their artists}
1. Releases > Artists.
2. Edit > Upload Avatar.
3. Submit [Expect] Success. \\
\midrule
M16 & \textbf{[title] Label user adds a new artist}
1. Releases > Artists.
2. Click Add New Artist.
3. Enter Name/Avatar.
4. Submit
5. View Artists [Expect] New Artist. \\
\midrule
M17 & \textbf{[title] User enters the song modification page without saving and returns to the homepage}
1. Releases > Edit song.
2. No changes made.
3. Click Return. \\
\midrule
M18 & \textbf{[title] Artist user logs in and changes avatar}
1. Change Avatar [Expect] Success. \\
\midrule
M19 & \textbf{[title] Creator joins WWA}
1. Studio > Work with Artists.
2. Apply [Expect] Apply Successfully. \\
\midrule
M20 & \textbf{[title] Creator submits video for WWA campaign}
1. Work with Artists > Select Campaign.
2. Click Use This Track.
3. Submit video [Expect] Success (status=200). \\
\midrule
M21 & \textbf{[title] Creator submits self-serve video}
1. Studio > Work with Artists.
2. Pass verify; Click Apply [Expect] Eligible. \\
\midrule
M22 & \textbf{[title] Creator enters WWA to view joined campaigns}
1. Studio > Work with Artists.
2. Tab: Personal Center.
3. Click Participated Campaigns [Expect] Review page. \\
\midrule
M23 & \textbf{[title] Creator enters WWA to favorite a campaign}
1. 1. Studio > Work with Artists > Campaign page.
2. Favorite campaign.
3. Check Personal Center [Expect] Success. \\
\midrule
M24 & \textbf{[title] Creator enters WWA to view bills}
1. Studio > Work with Artists.
2. Personal Center > Bills. \\
\midrule
M25 & \textbf{[title] Creator enters WWA to subscribe to a campaign}
1. Studio > Work with Artists.
2. Find Upcoming; Subscribe first one [Expect] Success. \\
\midrule
S1 & \textbf{[title] Unit conversion}
1. Search input: 'USD to CNY'. \\
\midrule
S2 & \textbf{[title] Game card}
1. Search input: 'Dota2'. \\
\bottomrule
\end{xltabular}
}

\bibliographystyle{ACM-Reference-Format}
\bibliography{main}

@ARTICLE{elbaum2008carving,
  author={Elbaum, Sebastian and Chin, Hui Nee and Dwyer, Matthew B. and Jorde, Matthew},
  journal={IEEE Transactions on Software Engineering}, 
  title={Carving and Replaying Differential Unit Test Cases from System Test Cases}, 
  year={2009},
  volume={35},
  number={1},
  pages={29-45},
  doi={10.1109/TSE.2008.103}}

@inproceedings{alshahwan2024observation,
author = {Alshahwan, Nadia and Harman, Mark and Marginean, Alexandru and Tal, Rotem and Wang, Eddy},
title = {Observation-Based Unit Test Generation at Meta},
year = {2024},
isbn = {9798400706585},
publisher = {Association for Computing Machinery},
address = {New York, NY, USA},
url = {https://doi.org/10.1145/3663529.3663838},
doi = {10.1145/3663529.3663838},
booktitle = {Companion Proceedings of the 32nd ACM International Conference on the Foundations of Software Engineering},
pages = {173–184},
numpages = {12},
location = {Porto de Galinhas, Brazil},
series = {FSE 2024}
}

@INPROCEEDINGS{gambi2023action,
  author={Gambi, Alessio and Gouni, Hemant and Berreiter, Daniel and Tymofyeyev, Vsevolod and Fazzini, Mattia},
  booktitle={2023 IEEE International Conference on Software Testing, Verification and Validation Workshops (ICSTW)}, 
  title={Action-Based Test Carving for Android Apps}, 
  year={2023},
  volume={},
  number={},
  pages={107-116},
  doi={10.1109/ICSTW58534.2023.00032}}

@misc{sathaye2023creating,
    title={Creating CLI packages and API playbooks from codified graphical user experience designs},
    author={Sathaye, Sumedh and East, Patrick and Kovetz, Reut and Minarik, Jennifer and Lisai, Kelly and Lent, Arthur and Reineke, Nicole},
    year={2023},
    month=aug # "~29",
    publisher={Google Patents},
    note={US Patent 11,740,878}
}

@INPROCEEDINGS{yandrapally2023carving,
  author={Yandrapally, Rahulkrishna and Sinha, Saurabh and Tzoref-Brill, Rachel and Mesbah, Ali},
  booktitle={2023 IEEE/ACM 45th International Conference on Software Engineering (ICSE)}, 
  title={Carving UI Tests to Generate API Tests and API Specification}, 
  year={2023},
  volume={},
  number={},
  pages={1971-1982},
  doi={10.1109/ICSE48619.2023.00167}}

@inproceedings{chen2024chatunitest,
author = {Chen, Yinghao and Hu, Zehao and Zhi, Chen and Han, Junxiao and Deng, Shuiguang and Yin, Jianwei},
title = {ChatUniTest: A Framework for LLM-Based Test Generation},
year = {2024},
isbn = {9798400706585},
publisher = {Association for Computing Machinery},
address = {New York, NY, USA},
url = {https://doi.org/10.1145/3663529.3663801},
doi = {10.1145/3663529.3663801},
booktitle = {Companion Proceedings of the 32nd ACM International Conference on the Foundations of Software Engineering},
pages = {572–576},
numpages = {5},
location = {Porto de Galinhas, Brazil},
series = {FSE 2024}
}

@inproceedings{sapozhnikov2024testspark,
author = {Sapozhnikov, Arkadii and Olsthoorn, Mitchell and Panichella, Annibale and Kovalenko, Vladimir and Derakhshanfar, Pouria},
title = {TestSpark: IntelliJ IDEA's Ultimate Test Generation Companion},
year = {2024},
isbn = {9798400705021},
publisher = {Association for Computing Machinery},
address = {New York, NY, USA},
url = {https://doi.org/10.1145/3639478.3640024},
doi = {10.1145/3639478.3640024},
booktitle = {Proceedings of the 2024 IEEE/ACM 46th International Conference on Software Engineering: Companion Proceedings},
pages = {30–34},
numpages = {5},
location = {Lisbon, Portugal},
series = {ICSE-Companion '24}
}

@ARTICLE{fakhoury2024llm,
  author={Fakhoury, Sarah and Naik, Aaditya and Sakkas, Georgios and Chakraborty, Saikat and Lahiri, Shuvendu K.},
  journal={IEEE Transactions on Software Engineering}, 
  title={LLM-Based Test-Driven Interactive Code Generation: User Study and Empirical Evaluation}, 
  year={2024},
  volume={50},
  number={9},
  pages={2254-2268},
  doi={10.1109/TSE.2024.3428972}}

@inproceedings{kim2024leveraging,
author = {Kim, Myeongsoo and Stennett, Tyler and Shah, Dhruv and Sinha, Saurabh and Orso, Alessandro},
title = {Leveraging Large Language Models to Improve REST API Testing},
year = {2024},
isbn = {9798400705007},
publisher = {Association for Computing Machinery},
address = {New York, NY, USA},
url = {https://doi.org/10.1145/3639476.3639769},
doi = {10.1145/3639476.3639769},
booktitle = {Proceedings of the 2024 ACM/IEEE 44th International Conference on Software Engineering: New Ideas and Emerging Results},
pages = {37–41},
numpages = {5},
location = {Lisbon, Portugal},
series = {ICSE-NIER'24}
}

@INPROCEEDINGS{deljouyi2023generating,
  author={Deljouyi, Amirhossein and Zaidman, Andy},
  booktitle={2023 IEEE 23rd International Working Conference on Source Code Analysis and Manipulation (SCAM)}, 
  title={Generating Understandable Unit Tests through End-to-End Test Scenario Carving}, 
  year={2023},
  volume={},
  number={},
  pages={107-118},
  doi={10.1109/SCAM59687.2023.00021},
}

@ARTICLE{wang2024software,
  author={Wang, Junjie and Huang, Yuchao and Chen, Chunyang and Liu, Zhe and Wang, Song and Wang, Qing},
  journal={IEEE Transactions on Software Engineering}, 
  title={Software Testing With Large Language Models: Survey, Landscape, and Vision}, 
  year={2024},
  volume={50},
  number={4},
  pages={911-936},
  doi={10.1109/TSE.2024.3368208}}

@article{guo2025comprehensive,
  title={A Comprehensive Survey on Benchmarks and Solutions in Software Engineering of LLM-Empowered Agentic System},
  author={Guo, Jiale and Huang, Suizhi and Li, Mei and Huang, Dong and Chen, Xingsheng and Zhang, Regina and Guo, Zhijiang and Yu, Han and Yiu, Siu-Ming and Lio, Pietro and others},
  journal={arXiv preprint arXiv:2510.09721},
  year={2025},
  doi={10.48550/arXiv.2510.09721}
}

@article{jin2024llms,
  title={From llms to llm-based agents for software engineering: A survey of current, challenges and future},
  author={Jin, Haolin and Huang, Linghan and Cai, Haipeng and Yan, Jun and Li, Bo and Chen, Huaming},
  journal={arXiv preprint arXiv:2408.02479},
  year={2024},
  doi={10.48550/arXiv.2408.02479}
}

@article{feng2025get,
  title={Get Experience from Practice: LLM Agents with Record \& Replay},
  author={Feng, Erhu and Zhou, Wenbo and Liu, Zibin and Chen, Le and Dong, Yunpeng and Zhang, Cheng and Zhao, Yisheng and Du, Dong and Hua, Zhichao and Xia, Yubin and others},
  journal={arXiv preprint arXiv:2505.17716},
  year={2025},
  doi={10.48550/arXiv.2505.17716}
}

@inproceedings{dinella2022toga,
author = {Dinella, Elizabeth and Ryan, Gabriel and Mytkowicz, Todd and Lahiri, Shuvendu K.},
title = {TOGA: a neural method for test oracle generation},
year = {2022},
isbn = {9781450392211},
publisher = {Association for Computing Machinery},
address = {New York, NY, USA},
url = {https://doi.org/10.1145/3510003.3510141},
doi = {10.1145/3510003.3510141},
booktitle = {Proceedings of the 44th International Conference on Software Engineering},
pages = {2130–2141},
numpages = {12},
location = {Pittsburgh, Pennsylvania},
series = {ICSE '22}
}

@article{molina2025test,
author = {Molina, Facundo and Gorla, Alessandra and d’Amorim, Marcelo},
title = {Test Oracle Automation in the Era of LLMs},
year = {2025},
issue_date = {June 2025},
publisher = {Association for Computing Machinery},
address = {New York, NY, USA},
volume = {34},
number = {5},
issn = {1049-331X},
url = {https://doi.org/10.1145/3715107},
doi = {10.1145/3715107},
journal = {ACM Trans. Softw. Eng. Methodol.},
month = may,
articleno = {150},
numpages = {24}
}

@article{zhang2025exploring,
author = {Zhang, Quanjun and Sun, Weifeng and Fang, Chunrong and Yu, Bowen and Li, Hongyan and Yan, Meng and Zhou, Jianyi and Chen, Zhenyu},
title = {Exploring Automated Assertion Generation via Large Language Models},
year = {2025},
issue_date = {March 2025},
publisher = {Association for Computing Machinery},
address = {New York, NY, USA},
volume = {34},
number = {3},
issn = {1049-331X},
url = {https://doi.org/10.1145/3699598},
doi = {10.1145/3699598},
journal = {ACM Trans. Softw. Eng. Methodol.},
month = feb,
articleno = {81},
numpages = {25}
}

@article{andrzejewski2025automated,
  title={Automated Test Generation Using Large Language Models},
  author={Andrzejewski, Marcin and Dubicka, Nina and Podolak, J{\k{e}}drzej and Kowal, Marek and Si{\l}ka, Jakub},
  journal={Data},
  volume={10},
  number={10},
  pages={156},
  year={2025},
  publisher={MDPI},
  doi={https://doi.org/10.3390/data10100156}
}

@inproceedings{ran2024guardian,
author = {Ran, Dezhi and Wang, Hao and Song, Zihe and Wu, Mengzhou and Cao, Yuan and Zhang, Ying and Yang, Wei and Xie, Tao},
title = {Guardian: A Runtime Framework for LLM-Based UI Exploration},
year = {2024},
isbn = {9798400706127},
publisher = {Association for Computing Machinery},
address = {New York, NY, USA},
url = {https://doi.org/10.1145/3650212.3680334},
doi = {10.1145/3650212.3680334},
booktitle = {Proceedings of the 33rd ACM SIGSOFT International Symposium on Software Testing and Analysis},
pages = {958–970},
numpages = {13},
location = {Vienna, Austria},
series = {ISSTA 2024}
}

@INPROCEEDINGS{mascia2025microservices,
  author={Mascia, Cristian and Guerriero, Antonio and Giamattei, Luca and Pietrantuono, Roberto and Russo, Stefano},
  booktitle={2025 IEEE/ACM Second International Conference on AI Foundation Models and Software Engineering (Forge)}, 
  title={Microservices Performance Testing with Causality-enhanced Large Language Models}, 
  year={2025},
  volume={},
  number={},
  pages={136-140},
  doi={10.1109/Forge66646.2025.00022}}

@inproceedings{harman2025mutation,
author = {Harman, Mark and Ritchey, Jillian and Harper, Inna and Sengupta, Shubho and Mao, Ke and Gulati, Abhishek and Foster, Christopher and Robert, Herv\'{e}},
title = {Mutation-Guided LLM-based Test Generation at Meta},
year = {2025},
isbn = {9798400712760},
publisher = {Association for Computing Machinery},
address = {New York, NY, USA},
url = {https://doi.org/10.1145/3696630.3728544},
doi = {10.1145/3696630.3728544},
booktitle = {Proceedings of the 33rd ACM International Conference on the Foundations of Software Engineering},
pages = {180–191},
numpages = {12},
location = {Clarion Hotel Trondheim, Trondheim, Norway},
series = {FSE Companion '25}
}

@inproceedings{li2022automatically,
author = {Li, Yinfeng and Gao, Chen and Du, Xiaoyi and Wei, Huazhou and Luo, Hengliang and Jin, Depeng and Li, Yong},
title = {Automatically Discovering User Consumption Intents in Meituan},
year = {2022},
isbn = {9781450393850},
publisher = {Association for Computing Machinery},
address = {New York, NY, USA},
url = {https://doi.org/10.1145/3534678.3539122},
doi = {10.1145/3534678.3539122},
booktitle = {Proceedings of the 28th ACM SIGKDD Conference on Knowledge Discovery and Data Mining},
pages = {3259–3269},
numpages = {11},
location = {Washington DC, USA},
series = {KDD '22}
}

@inproceedings{fraser2011evosuite,
author = {Fraser, Gordon and Arcuri, Andrea},
title = {EvoSuite: automatic test suite generation for object-oriented software},
year = {2011},
isbn = {9781450304436},
publisher = {Association for Computing Machinery},
address = {New York, NY, USA},
url = {https://doi.org/10.1145/2025113.2025179},
doi = {10.1145/2025113.2025179},
booktitle = {Proceedings of the 19th ACM SIGSOFT Symposium and the 13th European Conference on Foundations of Software Engineering},
pages = {416–419},
numpages = {4},
location = {Szeged, Hungary},
series = {ESEC/FSE '11}
}

@inproceedings{alshahwan2024automated,
author = {Alshahwan, Nadia and Chheda, Jubin and Finogenova, Anastasia and Gokkaya, Beliz and Harman, Mark and Harper, Inna and Marginean, Alexandru and Sengupta, Shubho and Wang, Eddy},
title = {Automated Unit Test Improvement using Large Language Models at Meta},
year = {2024},
isbn = {9798400706585},
publisher = {Association for Computing Machinery},
address = {New York, NY, USA},
url = {https://doi.org/10.1145/3663529.3663839},
doi = {10.1145/3663529.3663839},
booktitle = {Companion Proceedings of the 32nd ACM International Conference on the Foundations of Software Engineering},
pages = {185–196},
numpages = {12},
location = {Porto de Galinhas, Brazil},
series = {FSE 2024}
}

@inproceedings{alshahwan2024enhancing,
author = {Alshahwan, Nadia and Blasi, Arianna and Bojarczuk, Kinga and Ciancone, Andrea and Gucevska, Natalija and Harman, Mark and Krolikowski, Michal and Rojas, Rubmary and Martac, Dragos and Schellaert, Simon and Ustiuzhanina, Kate and Harper, Inna and Jia, Yue and Lewis, Will},
title = {Enhancing Testing at Meta with Rich-State Simulated Populations},
year = {2024},
isbn = {9798400705014},
publisher = {Association for Computing Machinery},
address = {New York, NY, USA},
url = {https://doi.org/10.1145/3639477.3639729},
doi = {10.1145/3639477.3639729},
booktitle = {Proceedings of the 46th International Conference on Software Engineering: Software Engineering in Practice},
pages = {1–12},
numpages = {12},
location = {Lisbon, Portugal},
series = {ICSE-SEIP '24}
}

@INPROCEEDINGS{beller2023learning,
  author={Beller, Moritz and Li, Hongyu and Nair, Vivek and Murali, Vijayaraghavan and Ahmad, Imad and Cito, Jürgen and Carlson, Drew and Aye, Ari and Dyer, Wes},
  booktitle={2023 IEEE/ACM International Conference on Automation of Software Test (AST)}, 
  title={Learning to Learn to Predict Performance Regressions in Production at Meta}, 
  year={2023},
  volume={},
  number={},
  pages={56-67},
  doi={10.1109/AST58925.2023.00010}}

@inproceedings {saokar2023servicerouter,
author = {Harshit Saokar and Soteris Demetriou and Nick Magerko and Max Kontorovich and Josh Kirstein and Margot Leibold and Dimitrios Skarlatos and Hitesh Khandelwal and Chunqiang Tang},
title = {{ServiceRouter}: Hyperscale and Minimal Cost Service Mesh at Meta},
booktitle = {17th USENIX Symposium on Operating Systems Design and Implementation (OSDI 23)},
year = {2023},
isbn = {978-1-939133-34-2},
address = {Boston, MA},
pages = {969--985},
url = {https://www.usenix.org/conference/osdi23/presentation/saokar},
publisher = {USENIX Association},
month = jul
}

@article{sun2023presto,
author = {Sun, Yutian and Meehan, Tim and Schlussel, Rebecca and Xie, Wenlei and Basmanova, Masha and Erling, Orri and Rosa, Andrii and Fan, Shixuan and Zhong, Rongrong and Thirupathi, Arun and Collooru, Nikhil and Wang, Ke and Agarwal, Sameer and Gupta, Arjun and Logothetis, Dionysios and Xirogiannopoulos, Kostas and Dutta, Amit and Gajjala, Varun and Jain, Rohit and Palakuzhy, Ajay and Pandian, Prithvi and Pershin, Sergey and Saikia, Abhisek and Shankhdhar, Pranjal and Somanchi, Neerad and Tailor, Swapnil and Tan, Jialiang and Viswanadha, Sreeni and Wen, Zac and Chattopadhyay, Biswapesh and Fan, Bin and Majeti, Deepak and Pandit, Aditi},
title = {Presto: A Decade of SQL Analytics at Meta},
year = {2023},
issue_date = {June 2023},
publisher = {Association for Computing Machinery},
address = {New York, NY, USA},
volume = {1},
number = {2},
url = {https://doi.org/10.1145/3589769},
doi = {10.1145/3589769},
journal = {Proc. ACM Manag. Data},
month = jun,
articleno = {189},
numpages = {25}
}

@INPROCEEDINGS{guo2025effectively,
  author={Guo, Wunan and Dong, Zhen and Shen, Liwei and Zhou, Daihong and Hu, Bin and Zhang, Chen and Xue, Hai},
  booktitle={2025 IEEE/ACM 33rd International Conference on Program Comprehension (ICPC)}, 
  title={Effectively Modeling UI Transition Graphs for Android Apps Via Reinforcement Learning}, 
  year={2025},
  volume={},
  number={},
  pages={13-24},
  doi={10.1109/ICPC66645.2025.00011}}

@inproceedings{cai2024reproducing,
author = {Cai, Xiaobao and Dong, Zhen and Wang, Yongjiang and Tiwari, Abhishek and Peng, Xin},
title = {Reproducing Timing-Dependent GUI Flaky Tests in Android Apps via a Single Event Delay},
year = {2024},
isbn = {9798400706127},
publisher = {Association for Computing Machinery},
address = {New York, NY, USA},
url = {https://doi.org/10.1145/3650212.3680377},
doi = {10.1145/3650212.3680377},
booktitle = {Proceedings of the 33rd ACM SIGSOFT International Symposium on Software Testing and Analysis},
pages = {1504–1515},
numpages = {12},
location = {Vienna, Austria},
series = {ISSTA 2024}
}

@inproceedings{zhang2024trace,
author = {Zhang, Chenxi and Dong, Zhen and Peng, Xin and Zhang, Bicheng and Chen, Miao},
title = {Trace-based Multi-Dimensional Root Cause Localization of Performance Issues in Microservice Systems},
year = {2024},
isbn = {9798400702174},
publisher = {Association for Computing Machinery},
address = {New York, NY, USA},
url = {https://doi.org/10.1145/3597503.3639088},
doi = {10.1145/3597503.3639088},
booktitle = {Proceedings of the IEEE/ACM 46th International Conference on Software Engineering},
articleno = {110},
numpages = {12},
location = {Lisbon, Portugal},
series = {ICSE '24}
}

@INPROCEEDINGS{chen2023dynamic,
  author={Chen, Yiru and Zhang, Chenxi and Dong, Zhen and Yang, Dingyu and Peng, Xin and Ou, Jiayu and Yang, Hong and Wu, Zheshun and Qu, Xiaojun and Li, Wei},
  booktitle={2023 38th IEEE/ACM International Conference on Automated Software Engineering (ASE)}, 
  title={Dynamic Graph Neural Networks-Based Alert Link Prediction for Online Service Systems}, 
  year={2023},
  volume={},
  number={},
  pages={79-90},
  doi={10.1109/ASE56229.2023.00177}}

@inproceedings{guo2022detecting,
author = {Guo, Wunan and Dong, Zhen and Shen, Liwei and Tian, Wei and Su, Ting and Peng, Xin},
title = {Detecting and fixing data loss issues in Android apps},
year = {2022},
isbn = {9781450393799},
publisher = {Association for Computing Machinery},
address = {New York, NY, USA},
url = {https://doi.org/10.1145/3533767.3534402},
doi = {10.1145/3533767.3534402},
booktitle = {Proceedings of the 31st ACM SIGSOFT International Symposium on Software Testing and Analysis},
pages = {605–616},
numpages = {12},
location = {Virtual, South Korea},
series = {ISSTA 2022}
}

@inproceedings{sun2021understanding,
author = {Sun, Jingling and Su, Ting and Li, Junxin and Dong, Zhen and Pu, Geguang and Xie, Tao and Su, Zhendong},
title = {Understanding and finding system setting-related defects in Android apps},
year = {2021},
isbn = {9781450384599},
publisher = {Association for Computing Machinery},
address = {New York, NY, USA},
url = {https://doi.org/10.1145/3460319.3464806},
doi = {10.1145/3460319.3464806},
booktitle = {Proceedings of the 30th ACM SIGSOFT International Symposium on Software Testing and Analysis},
pages = {204–215},
numpages = {12},
location = {Virtual, Denmark},
series = {ISSTA 2021}
}

@inproceedings{dong2020time,
author = {Dong, Zhen and B\"{o}hme, Marcel and Cojocaru, Lucia and Roychoudhury, Abhik},
title = {Time-travel testing of Android apps},
year = {2020},
isbn = {9781450371216},
publisher = {Association for Computing Machinery},
address = {New York, NY, USA},
url = {https://doi.org/10.1145/3377811.3380402},
doi = {10.1145/3377811.3380402},
booktitle = {Proceedings of the ACM/IEEE 42nd International Conference on Software Engineering},
pages = {481–492},
numpages = {12},
location = {Seoul, South Korea},
series = {ICSE '20}
}

\balance
\end{document}